\documentclass[reprint, 11pt, 3p,onehalfspacing, onecolumn]{elsarticle}
\usepackage{amsmath}
\usepackage{graphicx,subfig}
\usepackage{epsfig}
\usepackage{amssymb,amsthm,graphicx}
\usepackage{tabularx}
\usepackage{multirow}
\usepackage{tabularx}
\usepackage{float}
\usepackage{xcolor}
\usepackage{stackengine}
\usepackage{BOONDOX-cal}
\usepackage{mciteplus}
\usepackage{bm}
\biboptions{sort&compress}
\journal{Phys. Lett. A}
\begin{document}
	\begin{frontmatter}	
	\title{Dispersion Managed Generation of Peregrine Solitons and Kuznetsov-Ma Breather in an Optical Fiber} 	
	\author[DK]{Dipti Kanika Mahato}\ead{dkmahato@iitg.ac.in}
	\address[DK]{Department of Physics, Indian Institute of Technology  Guwahati, Guwahati-781039, Assam, India.}%
	\author[AG]{A. Govindarajan}
	\ead{govin.nld@gmail.com}
	\address[AG]{Centre for Nonlinear Dynamics, School of Physics, Bharathidasan University, Tiruchirappalli - 620 024, India}
	\author[AG]{M. Lakshmanan}\ead{lakshman.cnld@gmail.com}	\author[DK]{Amarendra K. Sarma\corref{Corresponding author}} 
	\ead{aksarma@iitg.ac.in}
	\cortext[Corresponding author]{Corresponding author}
	
		\begin{abstract}
		Optical rogue waves and its variants have been studied quite extensively in the context of optical fiber in recent years. It has been realized that dispersion management in optical fiber is experimentally much more feasible compared to its nonlinear counterpart. In this work, we report Kuznetsov-Ma (KM)-like breathers from the first three orders of rational solutions of the nonlinear Schr\"{o}dinger equation with periodic modulation of the dispersion coefficient along the fiber axis.  The breather dynamics are then controlled by proper choice of modulating parameters. Additionally, the evolution of new one-peak and two-peak breather-like solutions has been displayed corresponding to the second-order rational solution. Direct numerical simulations based on modulational instability has also been executed which agree well with the analytical results, thereby making the proposed system more feasible for experimental realization.
			\end{abstract}
		\begin{keyword}
			Peregrine solitons \sep periodic dispersion management \sep rogue waves \sep similarity transformation \sep pseudo-spectral method
		\end{keyword}
		
	\end{frontmatter}

	\section{\label{sec:level1}Introduction}
Since the publication of the remarkable study on optical rogue waves in 2007 \cite{solli2007optical} and later the experimental demonstration of Peregine solitons in nonlinear optical fiber \cite{kibler2010peregrine}, the study of optical rogue waves exploded in various contexts \cite{dudley2015rogue,bludov2013instabilities,onorato2013rogue,kibler2012observation,baronio2020resonant,xu2019breather,kraych2019nonlinear} in the last one decade. The theoretical understanding of optical rogue wave is provided by the so-called nonlinear Schr\"{o}dinger equation (NLSE).   
 
The study of NLSE has earned a lot of attention over the past few decades, due to its wide range of applications in numerous branches of science \cite{scott1982dynamics,scott1984magma,stenflo2010rogue}. Since its earliest usage in the study of deep water wave propagation in oceanography \cite{zakharov1968stability,lake1977nonlinear}, the self-focussing NLSE has been exploited extensively in different branches of physics, including nonlinear phenomena in optics \cite{haus1996solitons,kivshar2003optical}, Bose-Einstein condensate \cite{dalfovo1999pitaevskii,bludov2009matter,wen2011matter,rajendran2010,rajendran2011}, plasma physics \cite{shukla2010nonlinear,bailung2011observation,shukla2012alfvenic} and optomechanics \cite{gan2016solitons}. Here, the NLSE, which is under the investigation, refers to the self-focussing NLSE which has been used to study high amplitude extreme event phenomena such as rogue waves in optical fibers.
	
	The advantage of using the NLSE is that it belongs to a class of completely integrable nonlinear evolution equations and also admits an infinite number of exact solutions \cite{akhmediev1997solitons}. To elucidate, it exhibits envelope solitons on zero background \cite{shabat1972exact} and Kuznetsov-Ma (KM) breather \cite{ma1979perturbed}, Akhmediev breathers \cite{akhmediev1986modulation} and Peregrine solitons (PSs) \cite{peregrine1983water,cuevas2018stabilization} on finite background. Note that KM breathers are spatially localized solutions which breathe temporally while Akhmediev breathers are temporally localized solutions which breathe spatially and the Peregrine soliton is localized in both spatial and temporal co-ordinates. Each of these finite background solutions is specific limiting cases of a more general first-order two-parameter periodic solution with two periods, one along the spatial axis and the other along the temporal axis. A detailed relation between these first-order solutions corresponding to the lowest-order solutions of the NLSE can be found in Ref. \cite{akhmediev1997solitons}. The NLSE also admits another spatio-temporally localized solution as the limiting case of Akhmediev breather when its spatial period is considered to be infinite \cite{akhmediev1985generation,akhmediev1987exact}. This solution having the same expression as the Peregrine soliton, is known as the first-order rational soliton solution in the form of ratio of two polynomials both being functions of space  and time. Along similar lines, higher-order rational solutions were also obtained from the nonlinear superposition of multiple first-order rational solutions \cite{akhmediev2009extreme,akhmediev2009rogue,ankiewicz2011rogue,kedziora2012second,gaillard2013degenerate,kedziora2013classifying}.\
	
	The higher-order rational solutions of the constant coefficient NLSE, Peregrine soliton, Akhmediev breather and KM breather have been studied analytically in the context of rogue wave generation. It is important to note that a rogue wave is a spatiotemporally localized single wave with high amplitude which emerges suddenly and dies out rapidly without leaving any trace to follow. In case of an inhomogeneous Kerr nonlinear medium, such as optical fiber, the variation in system parameters (dispersion coefficient, nonlinearity and gain or loss) needs to be incorporated by employing the variable coefficient NLSE (Vc-NLSE). Similarity transformation has extensively been  used to construct analytical solutions of the Vc-NLSE connecting the solutions of constant coefficient NLSE \cite{kruglov2005exact}. Over the years utilizing this method, several theoretical studies have been reported on  rogue wave dynamics \cite{yan2010nonautonomous,zhong2013rogue,loomba2013optical,yang2018controllable,zhong2014controllable,li2018rogue}. Also, the study of breathers and rational solutions of different Vc-NLSE have revealed several new features, which include nonlinear tunnelling effect in periodically distributed system and exponentially dispersion decreasing fiber \cite{dai2012controllable} and Peregrine comb as multiple compression point in the amplitude in periodically modulated fibers \cite{tiofack2015comb}.

Recently it has been reported that periodic modulation of nonlinearity coefficient along the transverse axis  leads to evolution of Akhmediev-like breathers \cite{yang2018controllable}. Also, a previous report revealed how KM breathers and first and second order rational solutions evolve under periodic modulation of both dispersion and nonlinearity coefficients \cite{zhong2013rogue}. Zhong et al. have shown that under such condition, KM breathers propagate in a periodically modulated background with three peaks in one breathing unit, while the rational solutions maintain their standard features. In this work, by manipulating only the dispersion coefficient periodically along spatial axis, we obtain Kuznetsov-Ma (KM)-like breathers analytically from the first three orders of rational solutions of Vc-NLSE for the first time. The well-known similarity transformation has been used to solve the Vc-NLSE from the seed solutions of the standard NLSE. It should be noted that, we have chosen only rational solutions and observed their evolutions under periodic dispersion profile. Also, it has been observed that the evolution of such KM-like breathers does not depend on the order of the rational solutions. Additionally, two new different peak-dynamics corresponding to the second-order rational solution have been displayed by controlling the free parameters under the periodic dispersion profile. Also, direct numerical results have been shown by employing pseudo-spectral methods which confirm the analytical findings.
One of the major motivations for focussing on the dispersion management rather than the nonlinear one is due to the experimental feasibility of the earlier. While there are a number of experimental reports on successful modulation of dispersion management \cite{zhao2006gain,zhang2013experimental,chandrasekhar2006performance,torrengo2011experimental}, there is no experimental vindication of nonlinearity management in the context of nonlinear fiber optics albeit a few of the former were executed in Bose-Einstein condensate\cite{lin2011spin,huang2016experimental}.

The paper is organized as follows. Section 2 provides appropriate  model of (1+1)D Vc-NLSE with the analytical method to obtain the solutions. In Section 3, we explore rational solution dynamics and the evolution of the first, second and third-order rational solutions as controlled-breathers in the case of Vc-NLSE under periodic dispersion coefficient. Section 4 illustrates the numerical result. Finally, Section 5 provides an overall conclusion.

\section{\label{sec:level2}The Model and Similarity Transformation Technique}
The pulse propagation in a Kerr nonlinear medium, say an optical fiber, is best described by the so-called nonlinear Schr\"{o}dinger equation with variable coefficients, which reads as 
\begin{equation}
i\frac{\partial u}{\partial z} + \frac{\beta_2(z)}{2}\frac{\partial^2u}{\partial{x^2}} + \chi(z)|u|^2u= 0.
\end{equation}
where $u(z,x)$ denotes the complex envelope of the optical field, $z$ and $x$, respectively, are the propagation distance along the medium and the retarded time. Also, the parameters $\beta_2(z)$ and $\chi(z)$ are the group velocity dispersion (GVD) and the nonlinear (self-phase modulation) coefficients, respectively. Motivated by previous studies \cite{zhong2013rogue,loomba2013optical,yang2018controllable},  we utilize  the well-known similarity transformation in order to solve Eq. (1), which is written as
\begin{equation} 
u(z,x)=A(z)V(T,X)e^{iB(z,x)},
\end{equation}
As it is generally known, this similarity transformation reduces Eq. (1) to the known standard NLSE
\begin{equation} 
i\frac{\partial V}{\partial T} + \frac{1}{2}\frac{\partial^2V}{\partial{X^2}} + |V|^2V=0,
\end{equation}
where $V(T,X)$ represents the complex envelope of the optical field whose expression is known; $T(z)$ is the dimensionless propagation distance and $X(z,x)$ denotes the similarity variable which needs to be determined. Also, $A(z)$ and $B(z,x)$, both being real, are the amplitude and the phase function, respectively. Therefore, substitution of Eq. (2) into Eq. (1) will connect its solution to the exact known solution of Eq. (3), provided following set of relations and partial derivative equations (PDEs) are satisfied:
\begin{subequations}
	\begin{align}
	& AT_z=1,\\
	& \beta_2 A X_x^2=1,\\
	& \chi A^3=1,\\
	& B_z + \frac{\beta_2}{2}B_x^2=0,\\
	& \frac{A_z}{A} + \frac{\beta_2}{2}B_{xx}=0, \\
	& \frac{\beta_2 A}{2} X_{xx}=0,\\
	& B_x =- \frac{1}{\beta_2}\frac{X_z}{X_x}.
	\end{align}
\end{subequations}
In the above equations, subscripts denote partial derivatives with respect to $z$ or $x$. While solving Eqs. (4), it turns out that the variable coefficients of Eq. (1) spontaneously emerge in the parameters of similarity transformation as follows,
\begin{subequations}
	\begin{align}
	& T(z)=\int_{0}^{z}\frac{\beta_2(s)} {w^2(s)} ds, \\
	& w(z)= w_0\frac{\beta_2(z)}{\chi(z)},\\
	& X(z,x)=\frac{x}{w(z)}+\theta(z),\\
	& A(z)=\frac{1}{w(z)}\sqrt{\frac{\beta_2(z)}{\chi(z)}},\\
	& B(z,x)=\frac{w_z}{\beta_2 w}\frac{x^2}{2}+B_0(z),
	\end{align}
\end{subequations}
with a condition,
\begin{equation}
\beta_2 w_{zz}=\beta_{2z} w_z,
\end{equation} 
where $w(z)$ is the width of the rational soliton solution, $w_0$ being the initial width. Also, $\theta(z)$ and $B_0(z)$ are real integration constants which are chosen to be $\theta(z)=0$, $w_0=1$, $B_0(z)=1$ for the rest of the calculations. Finally, assembling all the solutions from Eqs. (5), the exact solution of Eq. (1) can now be obtained as given below
\begin{equation}
u_n(z,x)=\frac{1}{w(z)}\sqrt{\frac{\beta_2(z)}{\chi(z)}} V_n(T,X)e^{i\big(\frac{w_z}{\beta_2 w}\frac{x^2}{2} +\textcolor{red}{1}\big)}.
\end{equation}
where $n$ denotes the order of the solution.

As previously described, the NLSE (Eq. (3)) has several exact analytical solutions describing different physical phenomena. We will specifically deal with the first three orders of rational solutions obtained by the well-known Darboux transformation \cite{matveev1991darboux,cieslinski2009algebraic}. A well-established classification of these rational solutions is given by a complex parameter $s_j$, where $j$ is a positive integer, $j=1,2,...n$.

There is no $s_j$ parameter in the standard first-order rational solution, while there is one $s_1$ parameter in the second-order solution defined as $s_1=a+ib$. The second-order solution shows two different kinds of characteristics, type [0] when $s_1=0$ ($a=b=0$) and type [1] when $s_1\neq0$ \cite{akhmediev2011rogue}. Similarly, there exist two parameters  $s_1=a+ib$ and $s_2=c+id$ in the third-order rational solution. The third-order solutions are denoted as type [0,0] when $s_1=s_2=0$. For different $s_1$, $s_2$ values they are classified as type [0,1], [1,0] and [1,1] \cite{ling2013simple}. These parameters $a$, $b$, $c$ and $d$ are known as the `\emph{free parameters}'.On the basis of the work done in Refs.  \cite{ling2013simple,akhmediev2009rogue}, we use the first, second and the third-order rational solutions,
\begin{equation} 
V_1=\bigg[1-\frac{4(1+2iT)}{1+4X^2+4T^2}\bigg]e^{iT},
\end{equation}
\begin{equation} 
V_2=\bigg[1-2\frac{D_1}{D_2}\bigg]e^{iT},
\end{equation}
\begin{equation} 
V_3=\bigg[1-2\frac{\sum_{j=0}^{12} H_j X^{j}}{\sum_{j=0}^{12} F_j X^{j}}\bigg]e^{iT}
\end{equation}
and study the intensity distribution of Eq. (7) choosing a specific functional form of $\beta_2(z)$ and $\chi(z)$ for different $s_1$ and $s_2$ parameters. 
\\
\section{\label{sec:level4}Dynamics of rational solutions  for periodic dispersion profile}
In this section, we choose the dispersion coefficient to be periodic of the form, $\beta_2(z)=1+ \sigma$ cos $(\omega z)$. Here, $\sigma$ is the amplitude part of the modulation with $-1<\sigma<1$ and $\omega \neq 0$ is the spatial frequency. The corresponding pulse width, $w(z)$, the amplitude function, $A(z)$, and the similarity variable, $X(z,x)$, are given as follows:
$w(z)=\frac{1+\sigma \cos (\omega z)}{1+\sigma}$ , $A(z)=\frac{1+\sigma}{\sqrt{1+\sigma \cos (\omega z)}}$ , $X(z,x)=\frac{x(1+\sigma)}{1+\sigma \cos (\omega z)}$.We consider that there is no nonlinear management in the system, i.e., $\chi(z)=1$. It is worthwhile to note that such dispersion profiles could easily be generated in a fiber drawing process in experiments. In fact, numerous experimental and theoretical studies have been reported in the past where the dispersion coefficient is perturbed periodically, quasi-periodically or even randomly \cite{chertkov2001pulse,ablowitz2004dispersion,smith1996modulational,he2020dynamics,agrawal2013nonlinear,biswas2010mathematical}.

With these variables and Eqs. (7), (8), (9) and (12), we illustrate the intensity distribution of the first three orders of rational solutions for a periodically modulated group velocity dispersion.
\begin{figure}[ht]
	\centering
		\subfloat[]{\includegraphics[width=0.33\linewidth]{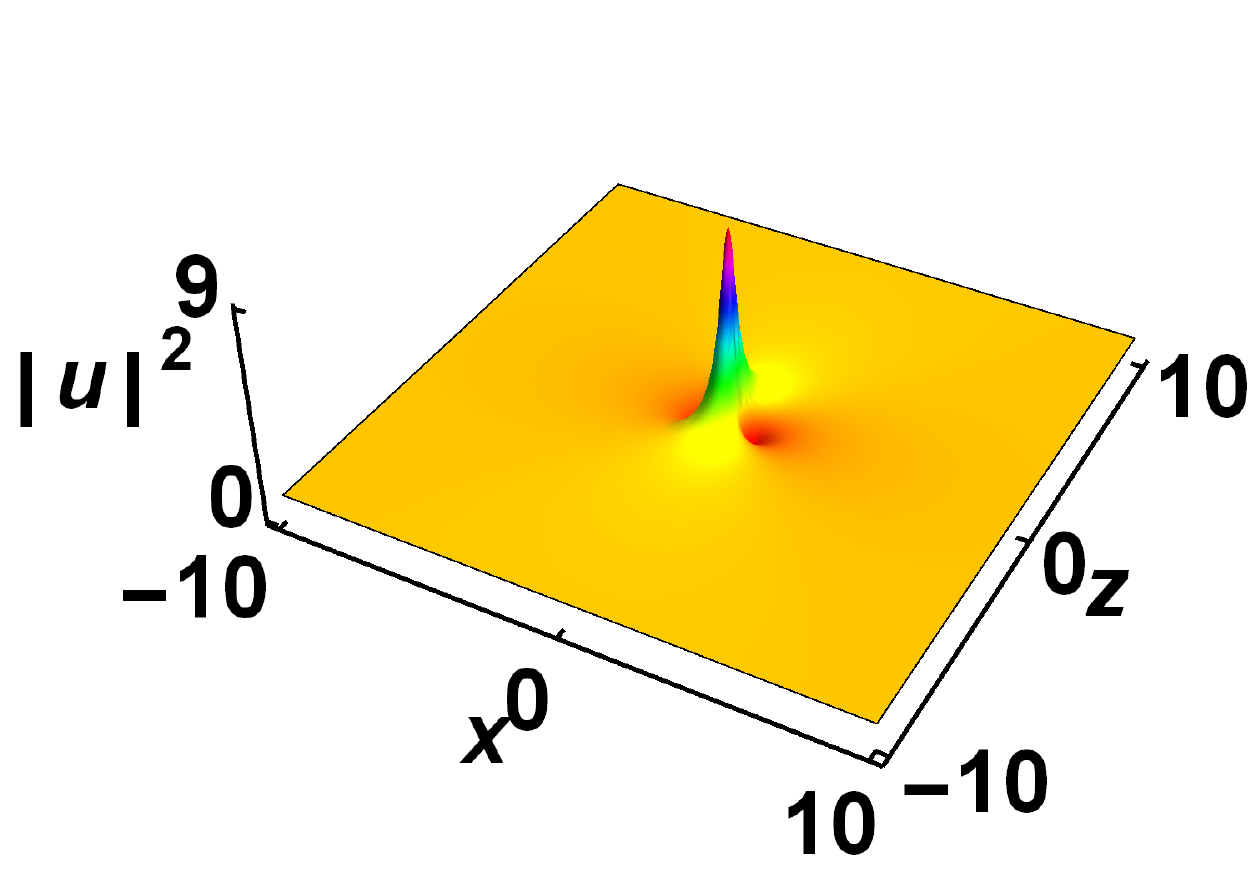}}
		\subfloat[]{\includegraphics[width=0.33\linewidth]{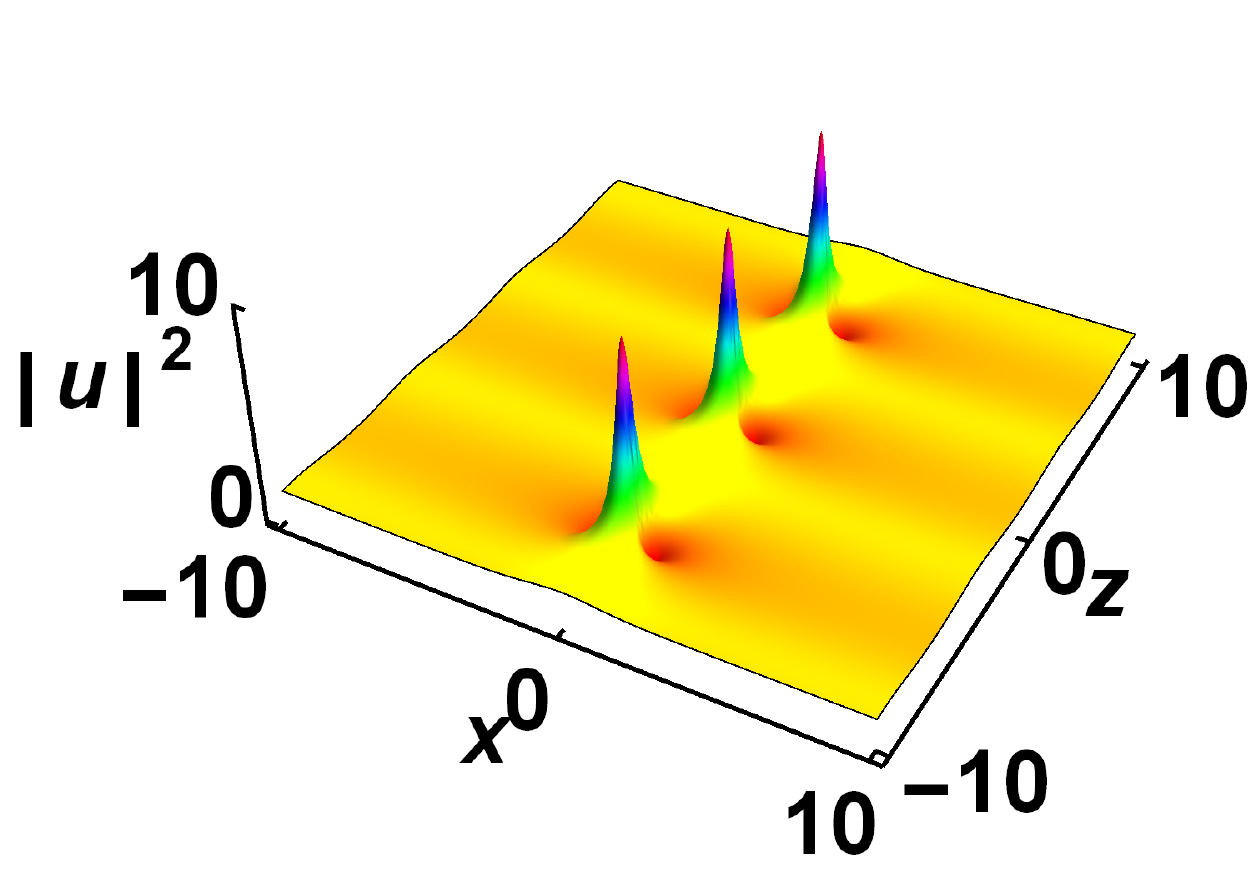}}\\
	\subfloat[]{\includegraphics[width=0.33\linewidth]{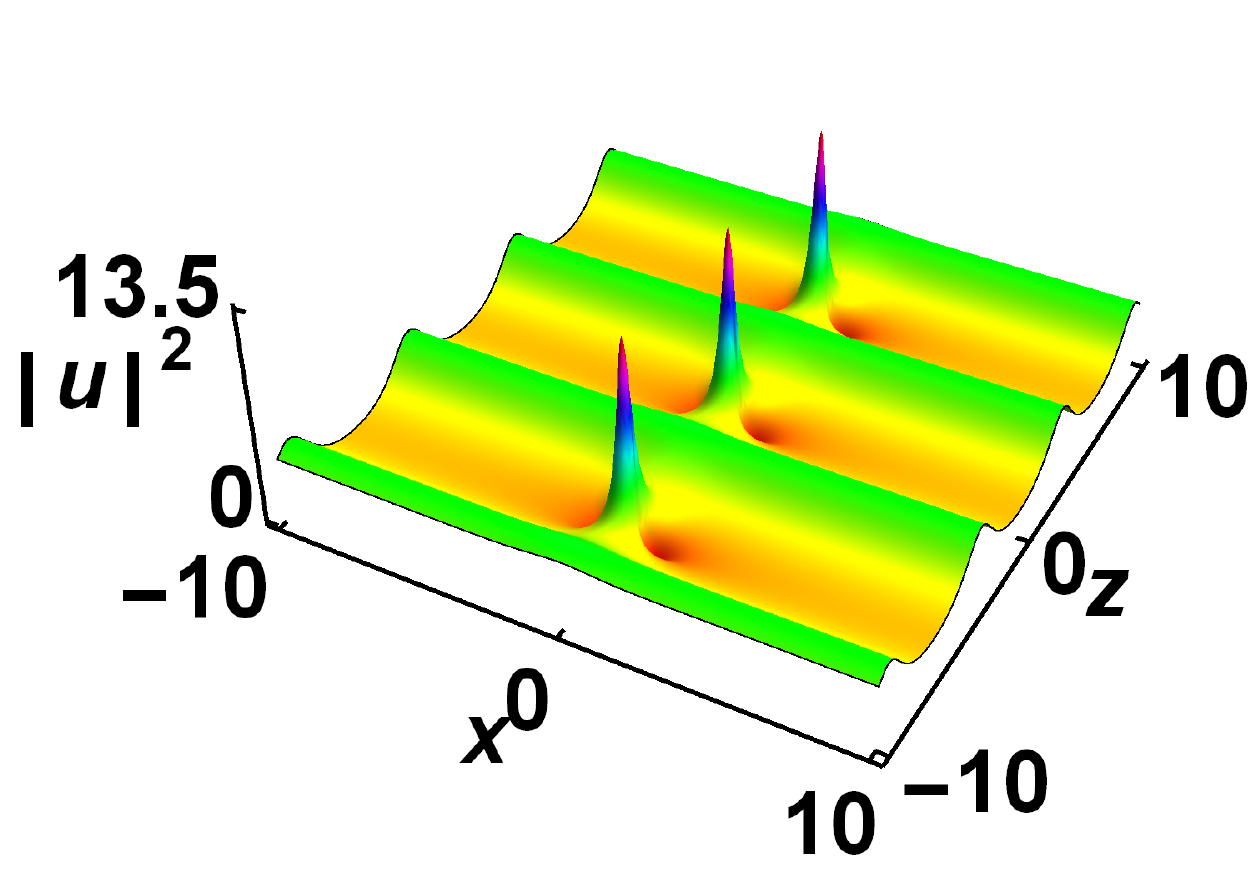}}
		\subfloat[]{\includegraphics[width=0.33\linewidth]{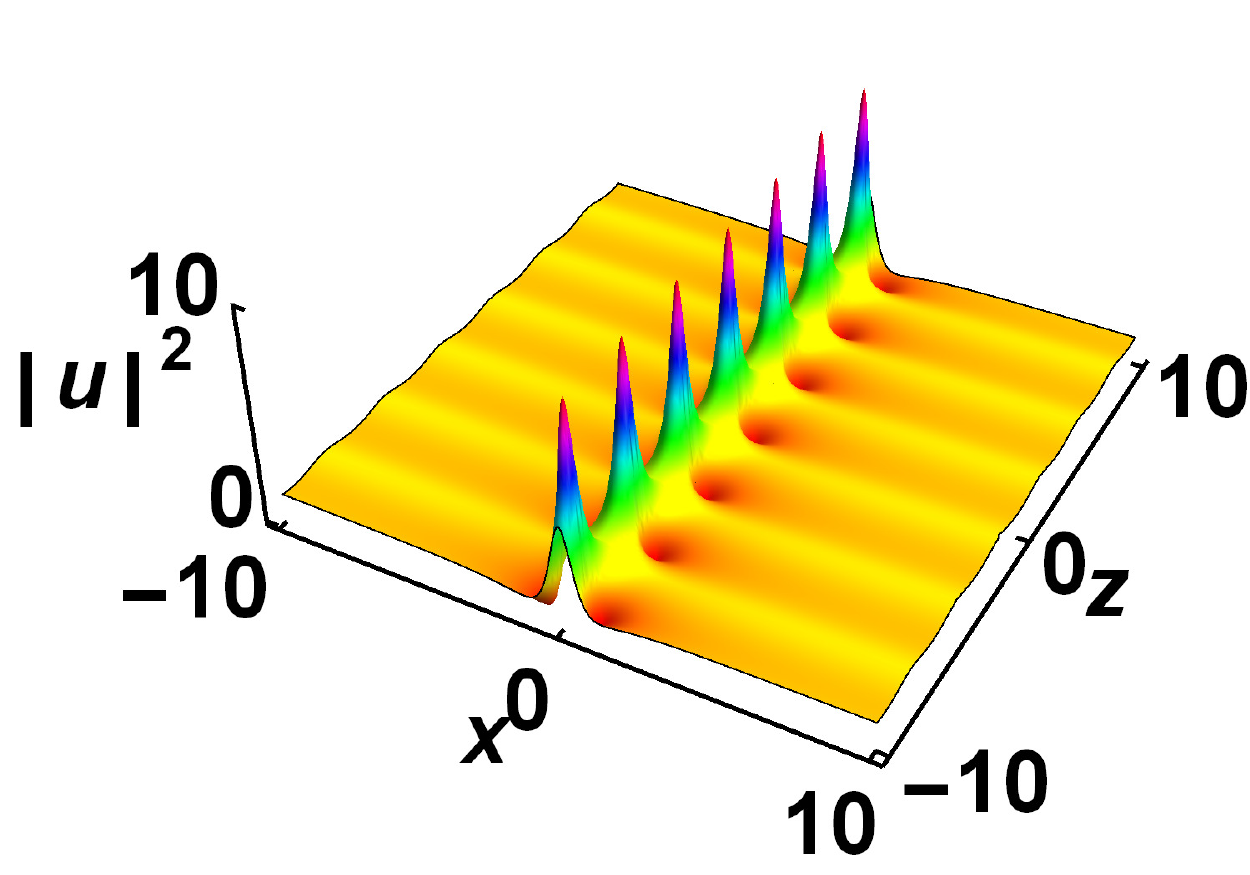}}
	\caption{Intensity distribution of the first-order rational solution (or Peregrine soliton). (a) Standard first-order solution without modulation. Controlled KM-like breather with: (b) $\omega=1$, $\sigma=0.1$; (c) $\omega=1$, $\sigma=0.5$ and (d) $\omega=2$, $\sigma=0.1$.}
	\label{fig: fig1}					
\end{figure}
It is well-known that the standard first-order rational solution (Peregrine soliton) shows a single peak, localized in both temporal and spatial axes (see Fig. 1(a)). Owing to the effect of periodic modulation of the dispersion along the propagation direction, the soliton solution starts breathing along the $z$-axis as shown in Fig. 1(b). Such a temporally breathing solution has strong resemblance with the KM breather, and thus it can be termed as `\textit{controlled KM-like breather}'. Here, the breather and the background have different frequencies in $z$, thus producing a beating between the breather solution and the background. The parameter, $\sigma$, controls the background amplitude and the breather peak power see Figs. 1(b) and 1(c) , while the spatial frequency ($\omega$) controls the breathing frequency of the KM-like breather and the periodicity of the finite background as shown in Fig. 1(d).
\begin{figure}[ht]
	\centering
	\subfloat[]{\includegraphics[width=0.33\linewidth]{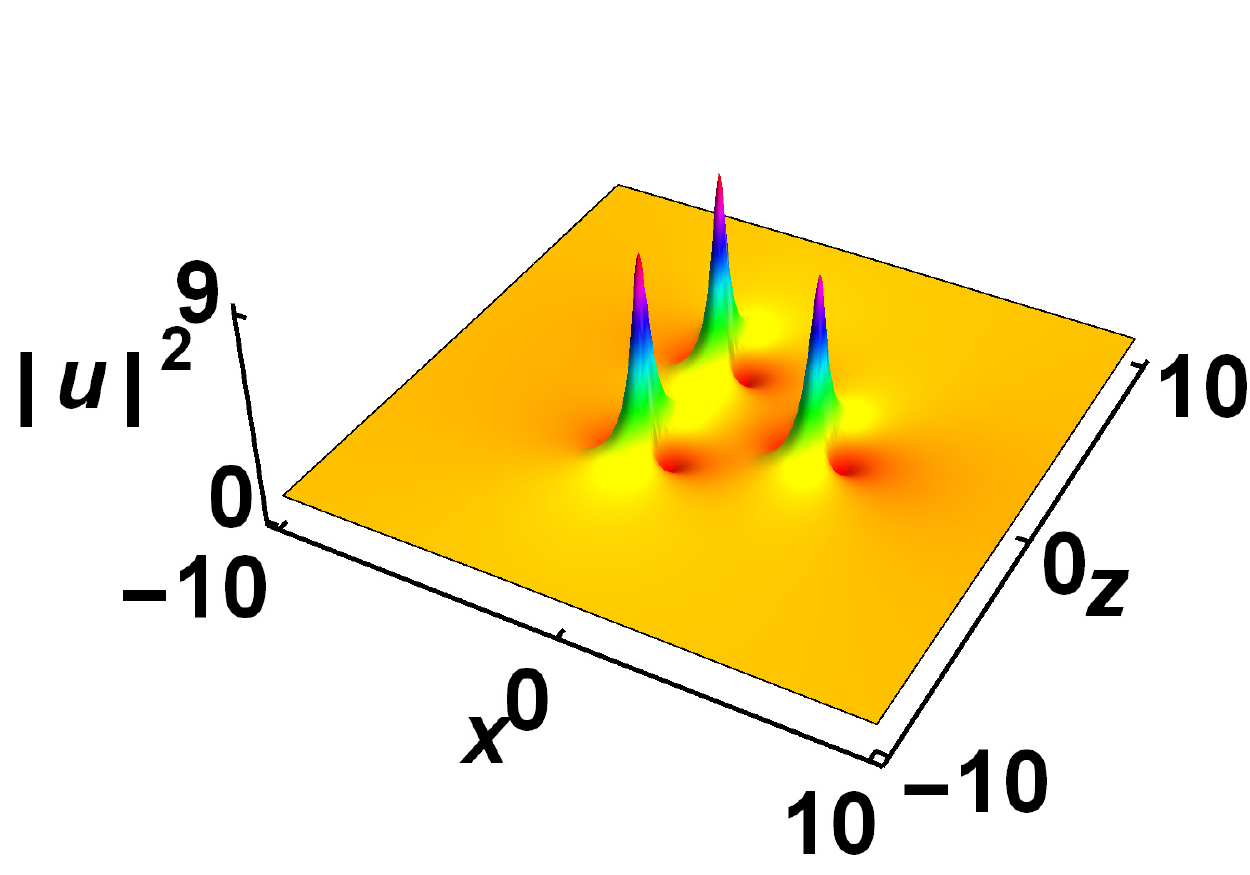}}
	\subfloat[]{\includegraphics[width=0.33\linewidth]{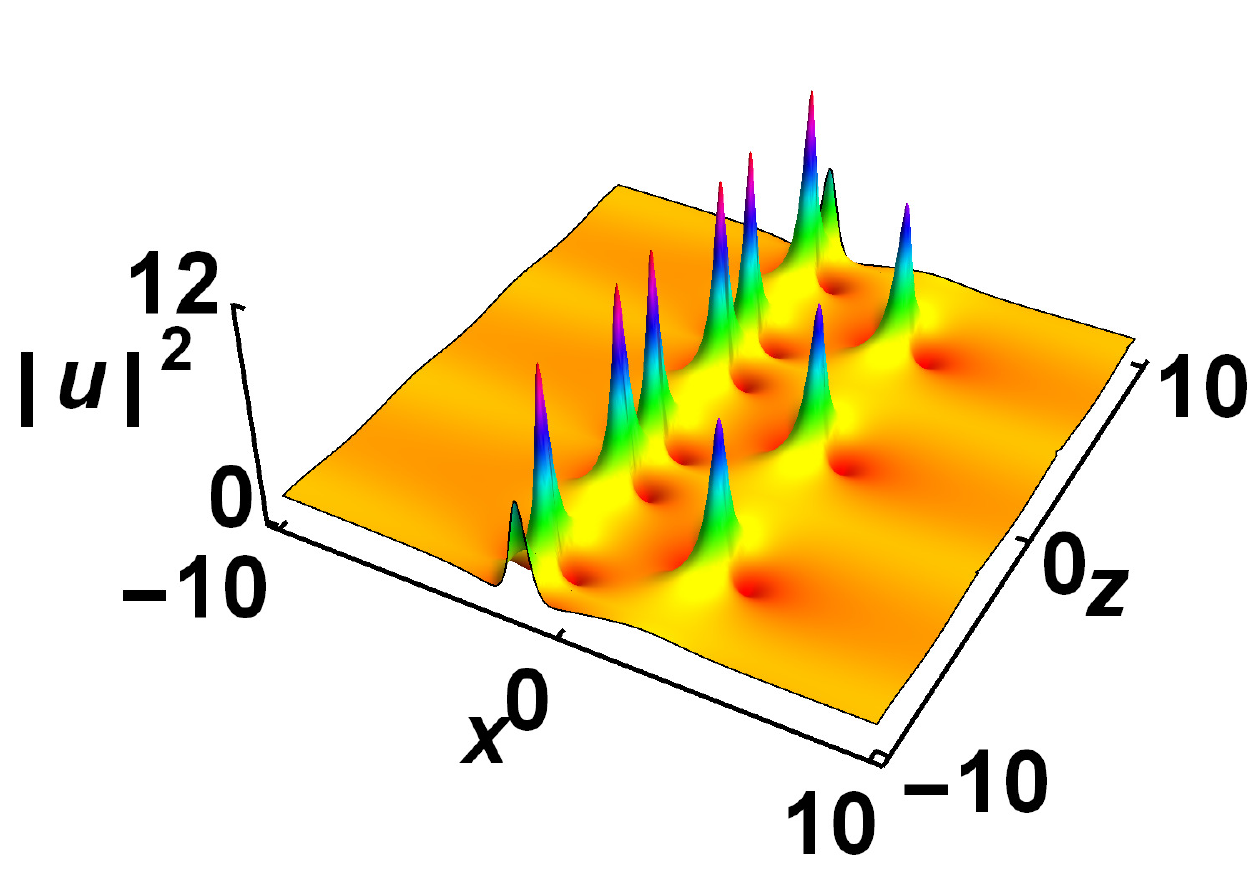}}\\
	\subfloat[]{\includegraphics[width=0.33\linewidth]{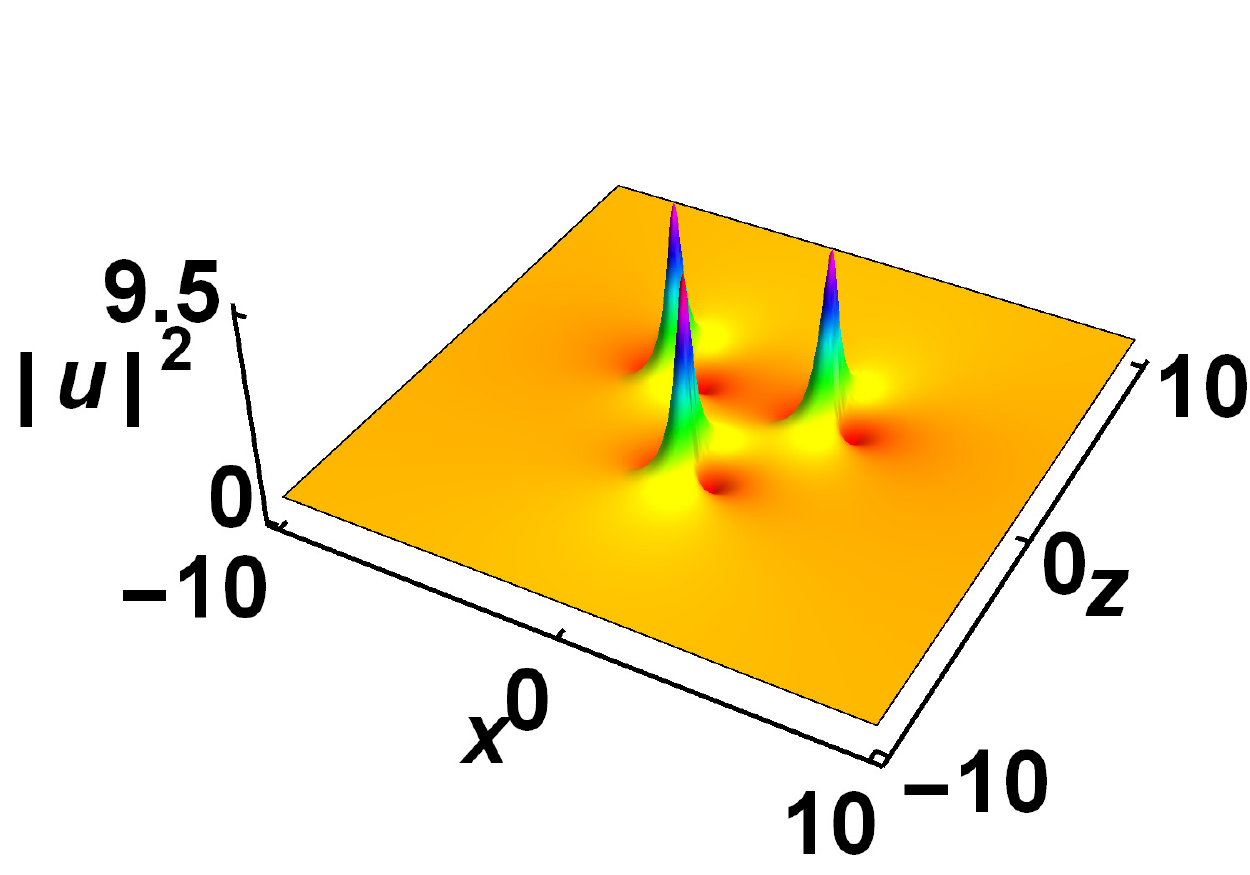}}
	\subfloat[]{\includegraphics[width=0.33\linewidth]{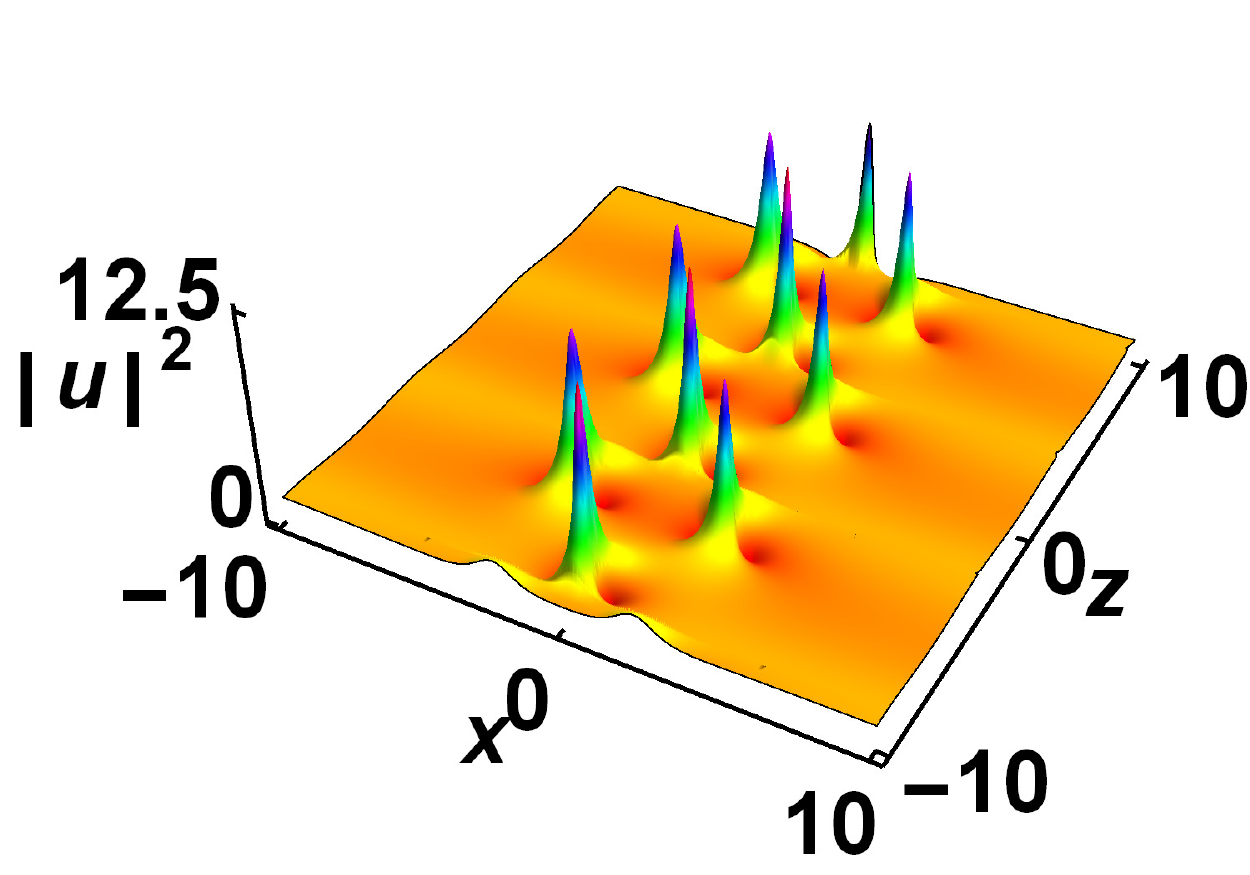}}
	\caption{Intensity distribution of the second-order rational solution. For $a=20$: (a) Standard second-order type [1] solution and (b) Controlled type [1] breather, both symmetrical about x-axis. For $b=20$: (c) Standard second-order type [1] solution, (d) Controlled type [1] breather, both symmetrical about z-axis. Modulating parameters chosen: $\sigma=0.1$, $\omega=1$.}
	\label{fig: fig2}					
\end{figure}

Similar controllable KM-like breather can be obtained in the case of second-order rational solution as well. As per the previously mentioned classification of the standard second-order rational solution which is based on $s_1$ parameter \cite{ling2013simple}, type [0] solution possesses single peak. On the other hand, type [1] solution possesses three peaks (triplets) distributed in triangular shape. These triplets in second-order solution are symmetric about the $x$-axis when $a\neq0$, $b=0$ as observed in Fig. 2(a) and about the $z$-axis when $a=0$, $b\neq0$. Considering specific values of $\sigma=0.1$ and $\omega=1$, it has been observed that both type [0] and type [1] solutions show breathing features under periodic dispersion. Here, we have shown the intensity evolution and contour plot of modulated type [1] solution in Figs. 2(c) and 2(d). This kind of solution could be termed as `\textit{controlled type [1] breather}'. Considering the triangular peak distribution in Figs. 2(a) as one unit, in each such periodic unit of the type [1] breather, the triplets are symmetric about the $x$-axis when $a\neq0$, $b=0$ (see Fig. 2(b)). On the other hand, the triplets are symmetric about the $z$-axis when $a=0$, $b\neq0$. The distance between the peaks gets increased with the increase in the value of either $a$ or $b$. Thus, type [1] breather maintains the characteristics of type [1] rational solution while breathing.
\begin{figure}[!ht]
	\centering
		\subfloat[]{\includegraphics[width=0.3\linewidth]{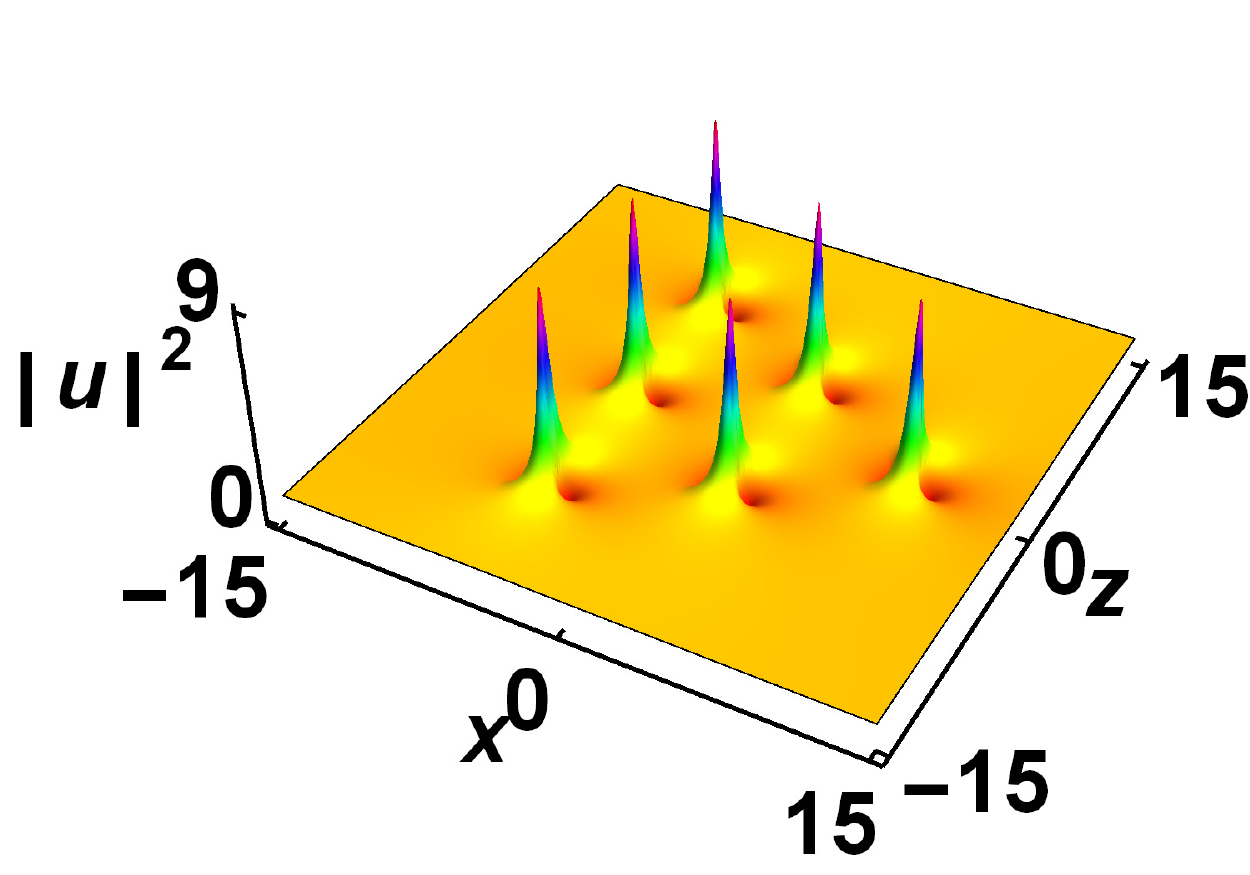}}
		\subfloat[]{\includegraphics[width=0.3\linewidth]{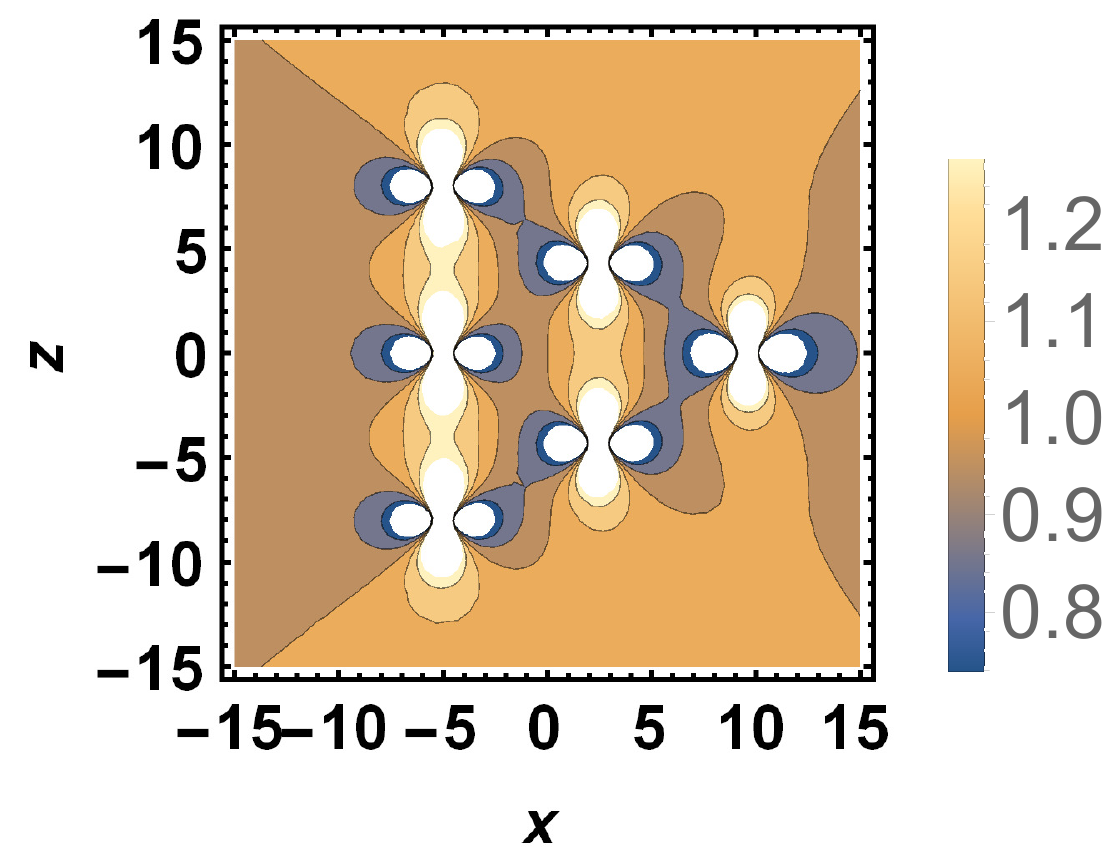}}\\
		\subfloat[]{\includegraphics[width=0.3\linewidth]{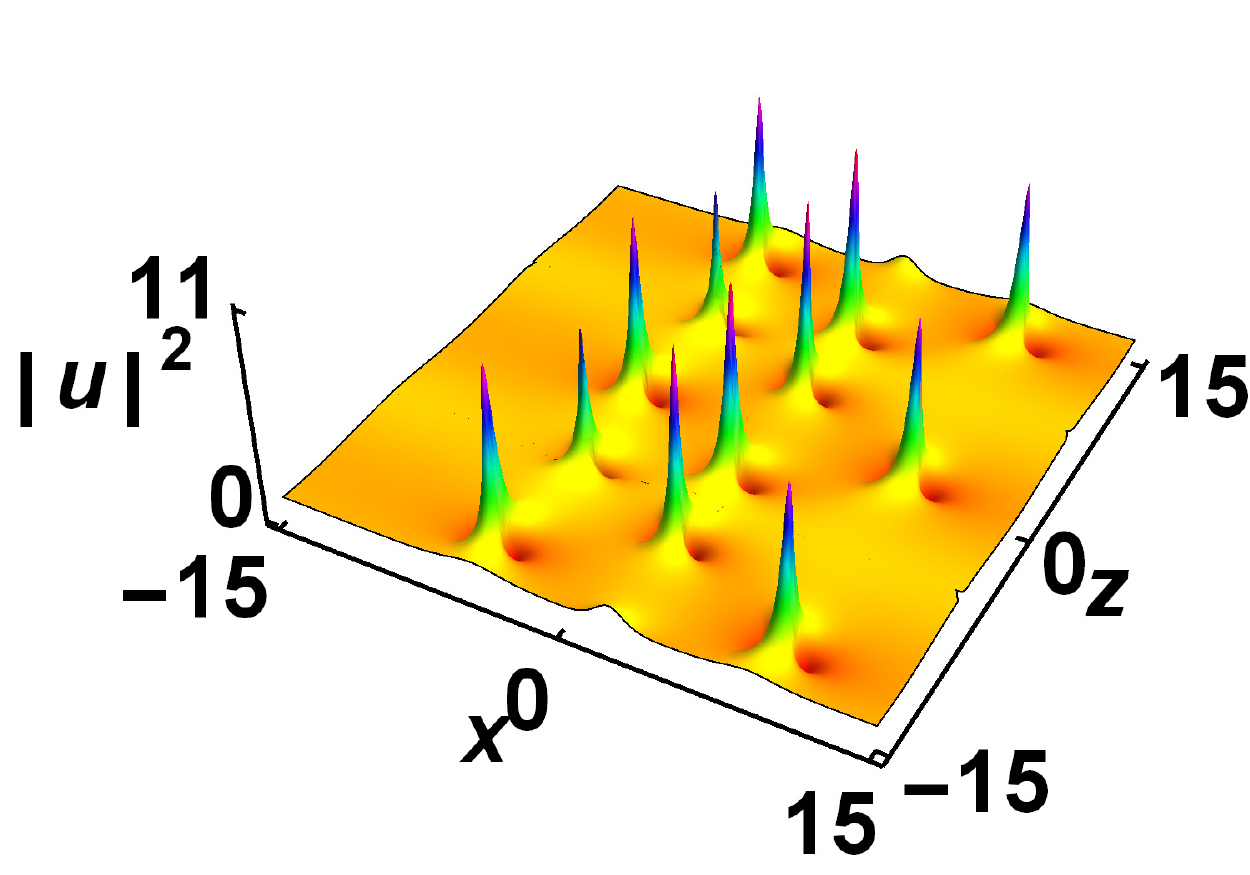}}
		\subfloat[]{\includegraphics[width=0.3\linewidth]{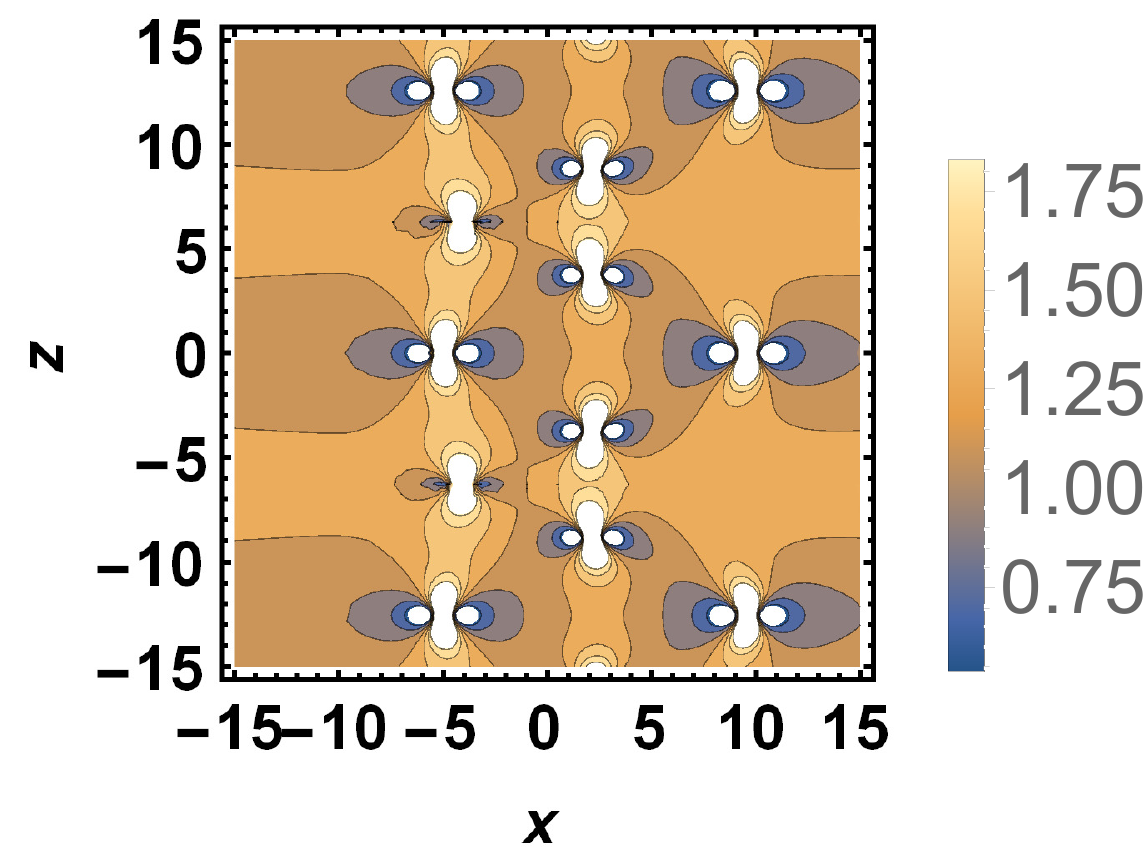}}\\
		\subfloat[]{\includegraphics[width=0.3\linewidth]{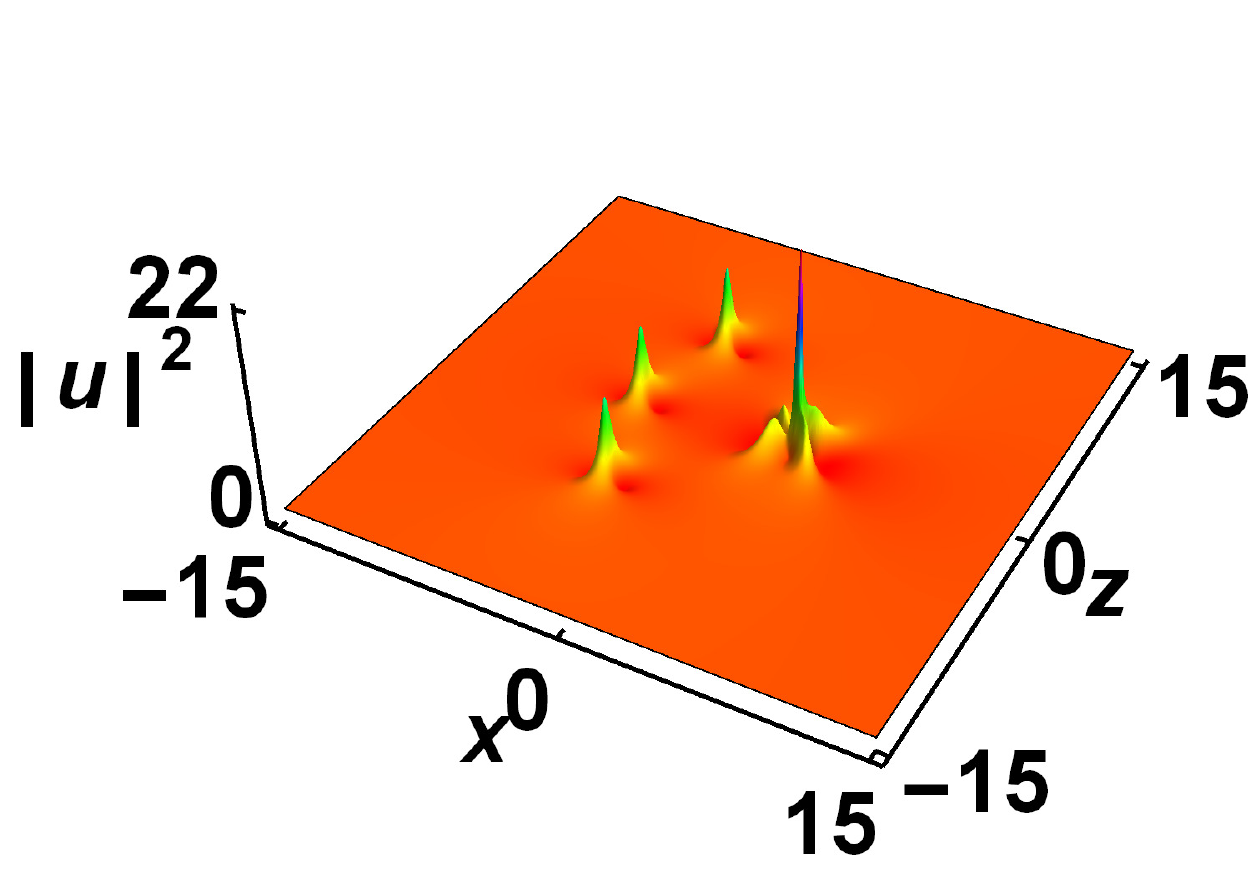}}
		\subfloat[]{\includegraphics[width=0.3\linewidth]{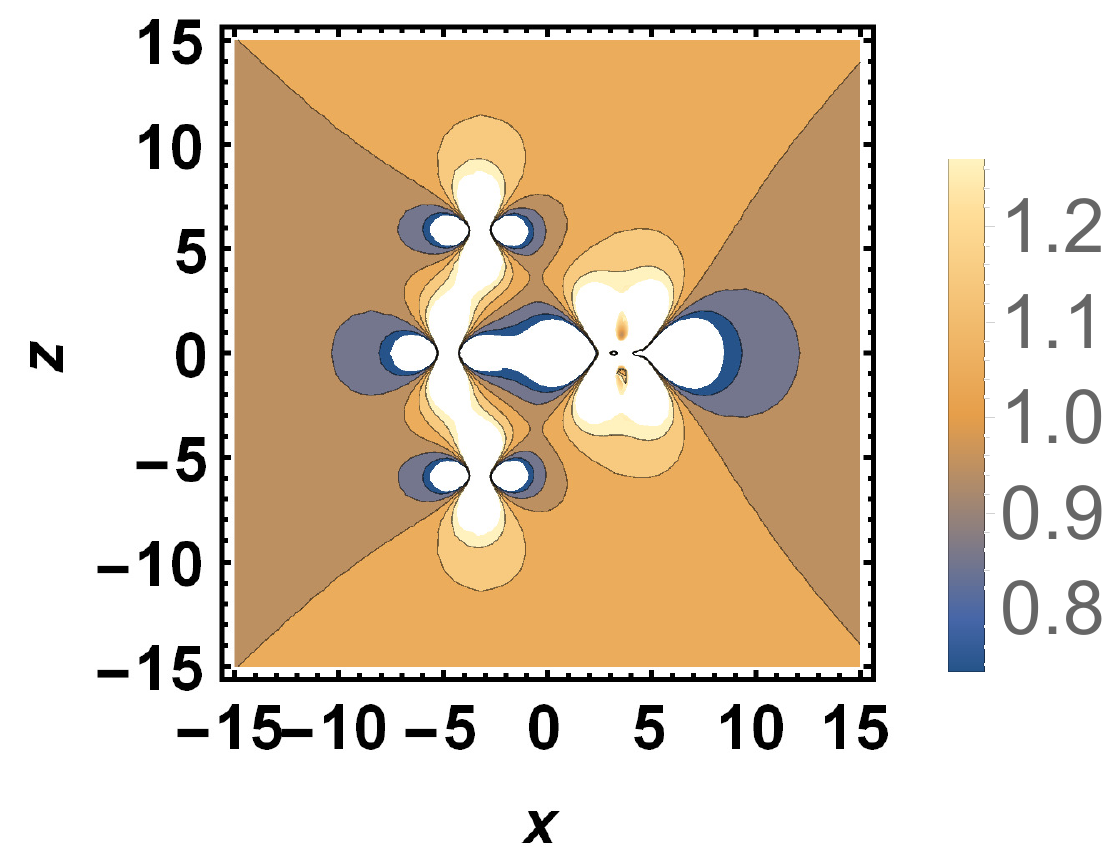}}\\
		\subfloat[]{\includegraphics[width=0.3\linewidth]{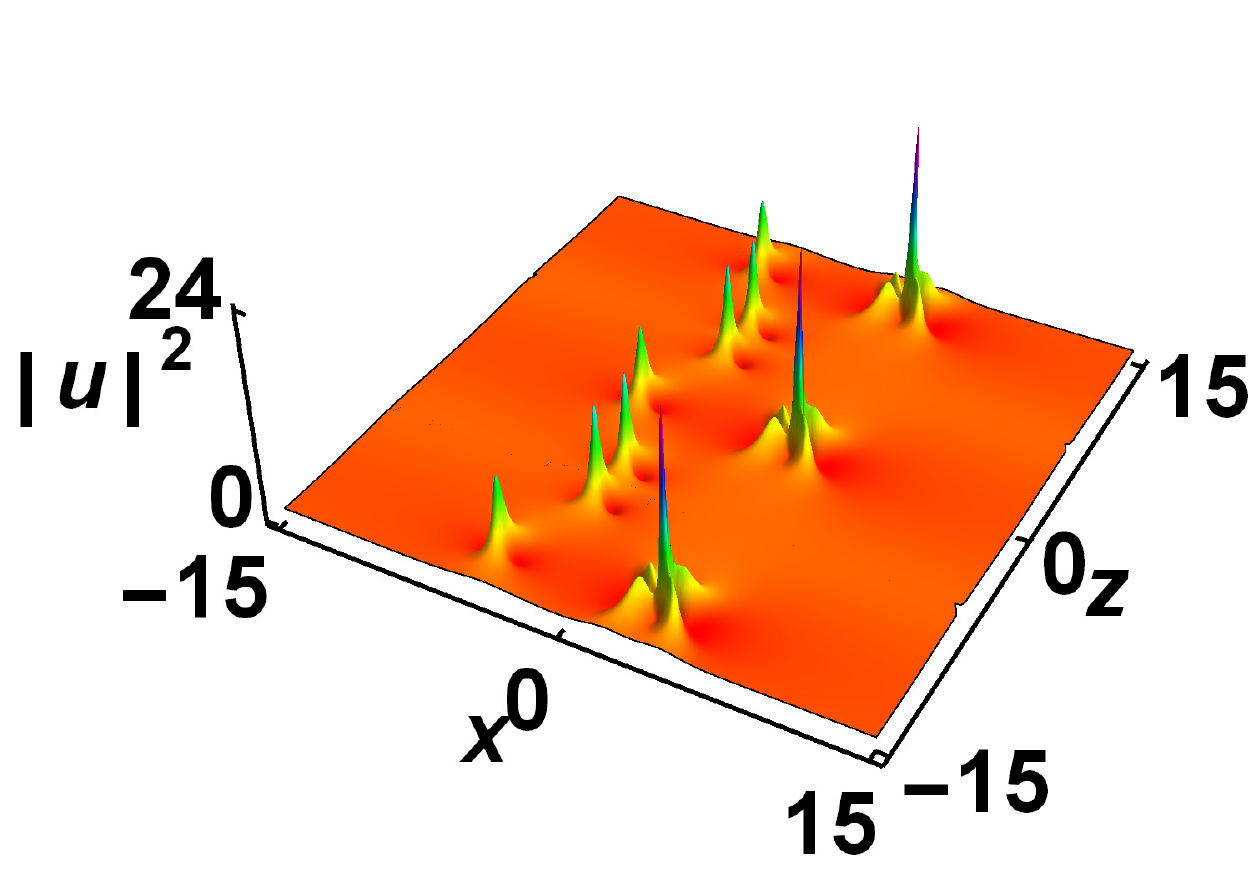}}
		\subfloat[]{\includegraphics[width=0.3\linewidth]{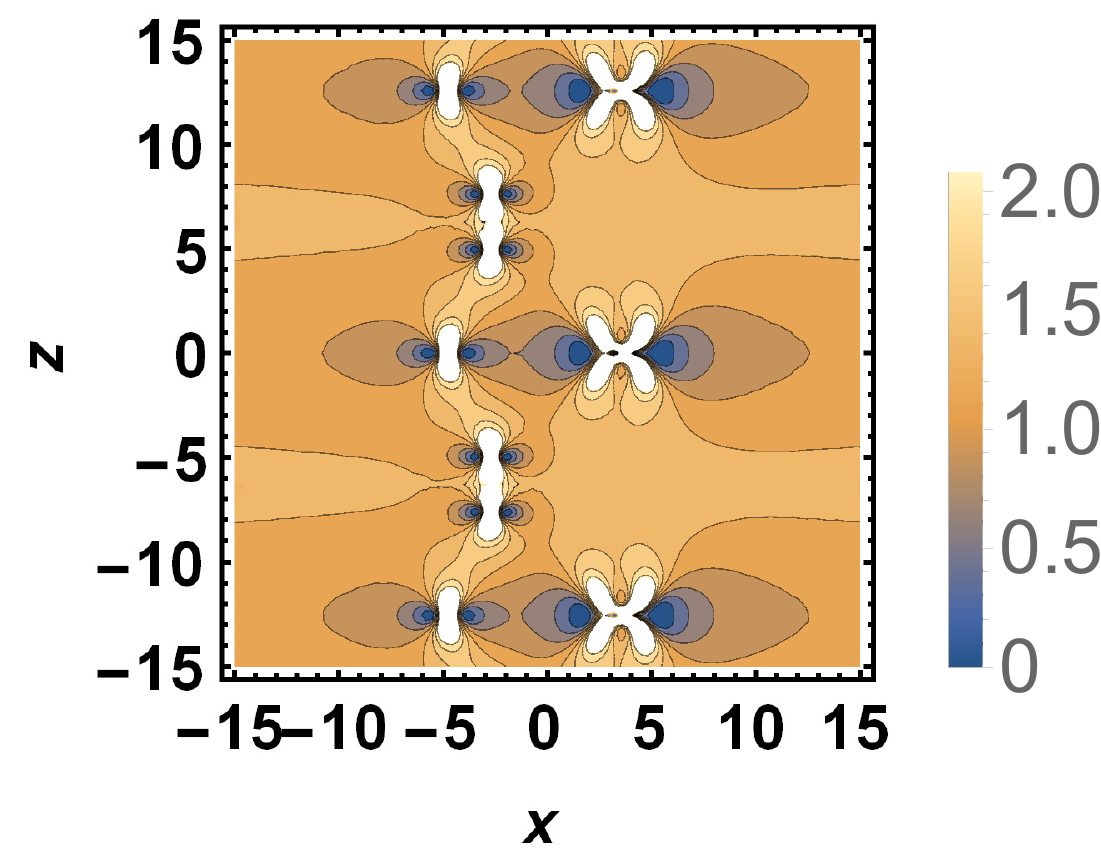}}
		\caption{Intensity distribution and contour plots for third-order rational solution. For $s_1=100$, $s_2=0$: (a),(b) Triangular cascade structure of the standard third-order type [1,0] solution and (c),(d) Controlled type [1,0] breather. For $s_1=31$, $s_2=500$: (e),(f) Claw-like structure of the standard third-order type [1,1] solution and (g),(h) Controlled type [1,1] breather. Modulating parameters chosen: $\sigma=0.1$, $\omega=0.5$.}
		\label{fig: fig3}						
\end{figure}

Unlike the first and the second-order rational solutions, the characterization of the basic third-order rational solution requires two parameters, $s_1$ and $s_2$ as already pointed out. From the previous researches, we know that the standard third-order solution shows four different features, namely, the one peak or type [0,0] \cite{akhmediev2009rogue}, the triangular cascade or type [1,0] \cite{guo2012nonlinear,kedziora2013classifying}, the pentagram structure or type [0,1] \cite{guo2012nonlinear,kedziora2013classifying} and the claw-structure or type [1,1] \cite{kedziora2012second,ling2013simple}. The intensity distribution of the standard third-order type [1,0] and type [1,1] solutions are shown in Fig. 3(a) with $s_1=100$, $s_2=0$ ($a=100$, $b=0$, $c=d=0$) and in Fig. 3(e) with $s_1=31$, $s_2=500$ ($a=31$, $b=0$, $c=500$, $d=0$), respectively. The standard type [1,0] solution corresponds to a unit of triangular structure, which is basically six first-order rational solutions of the same amplitude arranged in a triangular shape [Fig. 3(a)]. 

On the other hand, intensity distribution of type [1,1] solution displays clawlike structure which contains three first-order and one second-order solutions with their highest amplitude [Fig. 3(e)].
Now, considering $\sigma=0.1$ and $\omega=0.5$, it has been observed that, for $s_1=100$, $s_2=0$, the unit of triangular structure is getting repeated periodically along the $z$-axis [Fig. 3(c)]. Here, the extreme-end peaks of two consecutive units (at negative $x$-axis) combine together and appear with lower amplitude in comparison with the other peaks. For better understanding, the corresponding contour plot of controlled type [1,0] breather has been displayed in Fig. 3(d). For the claw-like structure, considering $s_1=31$, $s_2=500$ and $\sigma=0.1$, $\omega=0.5$, the periodic claw-like structure type [1,1] breather is observed under periodic modulation [Figs. 3(g) and 3(h)].
It can be observed that for specific values of modulating parameters $\sigma$ and $\omega$, we get KM-like breathers in the first three orders of rational solutions. Thus, we can interpret that it is also possible to generate and control KM-like breathers from higher-order rational solutions with simple periodic dispersion profile.

\subsection{\label{sec:level6}Three-peak Breather to Single-peak and Two-peak Breather conversion}
In this section, we briefly explore some unique aspects of the second-order rational solution for periodically varying dispersion profile. In the case of the constant coefficient NLSE solution, for $s_1\neq0$, if the value of real or imaginary part of $s_1$ parameter ($a$ or $b$) is increased, the peak separation gets increased for both the cases \cite{ling2013simple}. But under periodically varying dispersion profile, for a fixed value of the modulating parameters $\sigma$ and $\omega$, the breather dynamics show a completely different characteristics depending on the value of the real or imaginary part of $s_1$ parameter.
\begin{figure}[ht]
		\centering
		\subfloat[]{\includegraphics[width=0.33\linewidth]{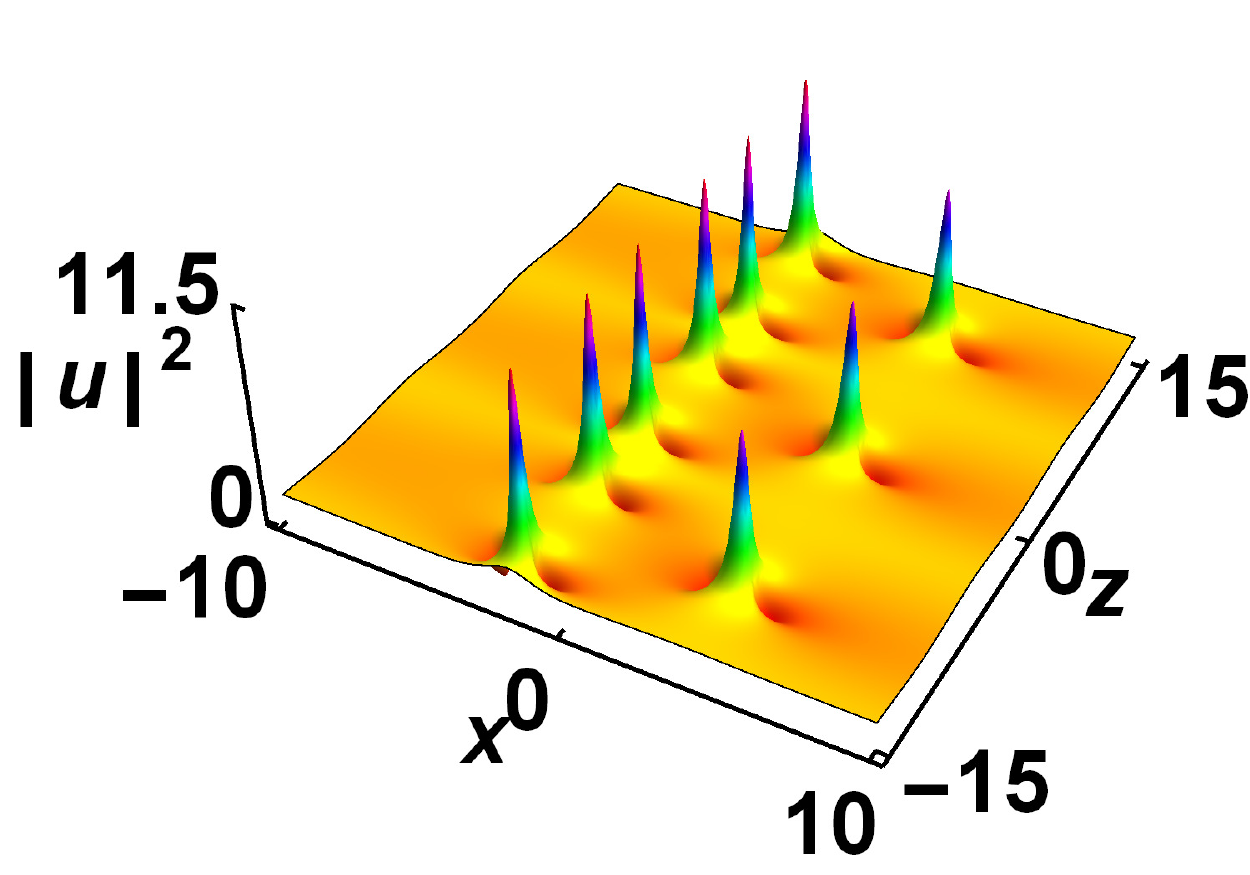}}
		\subfloat[]{\includegraphics[width=0.33\linewidth]{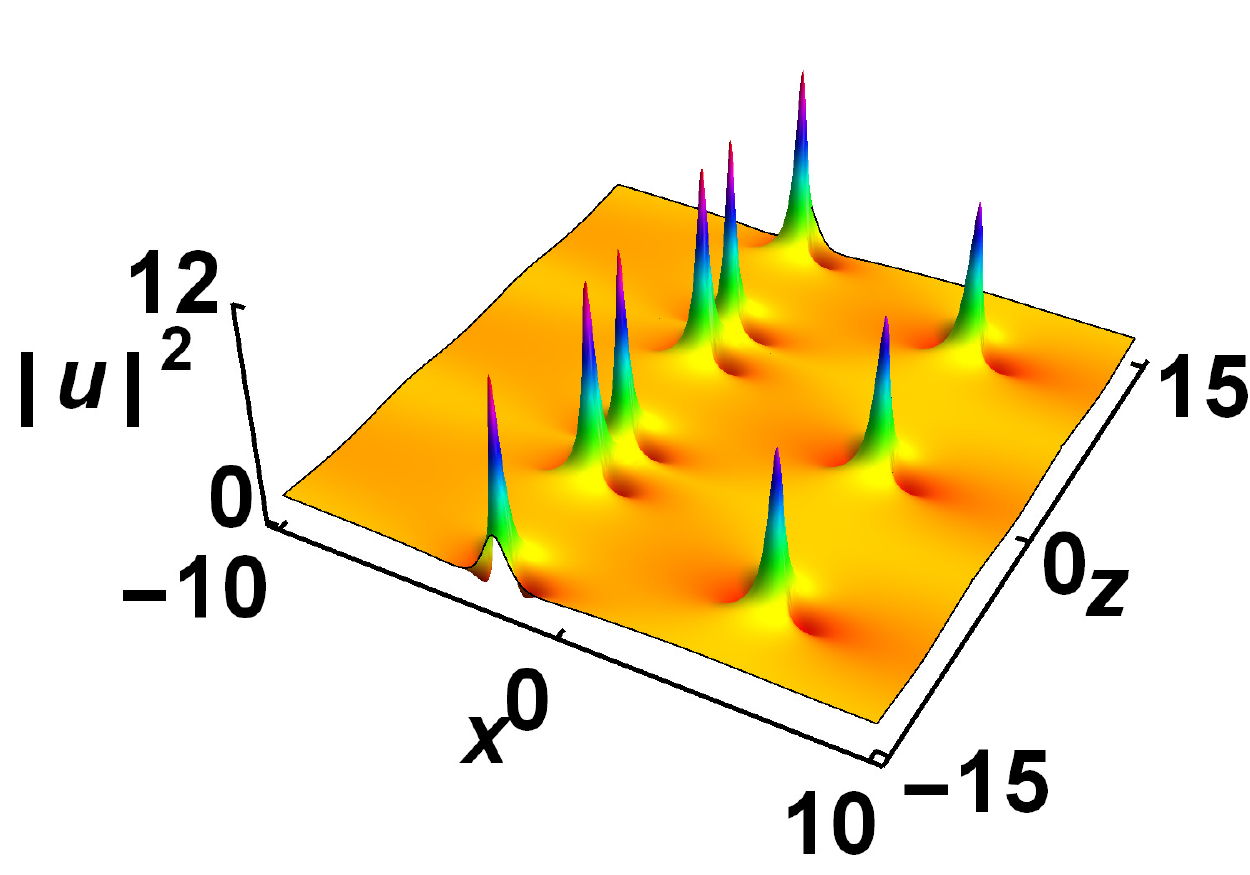}}
		\subfloat[]{\includegraphics[width=0.33\linewidth]{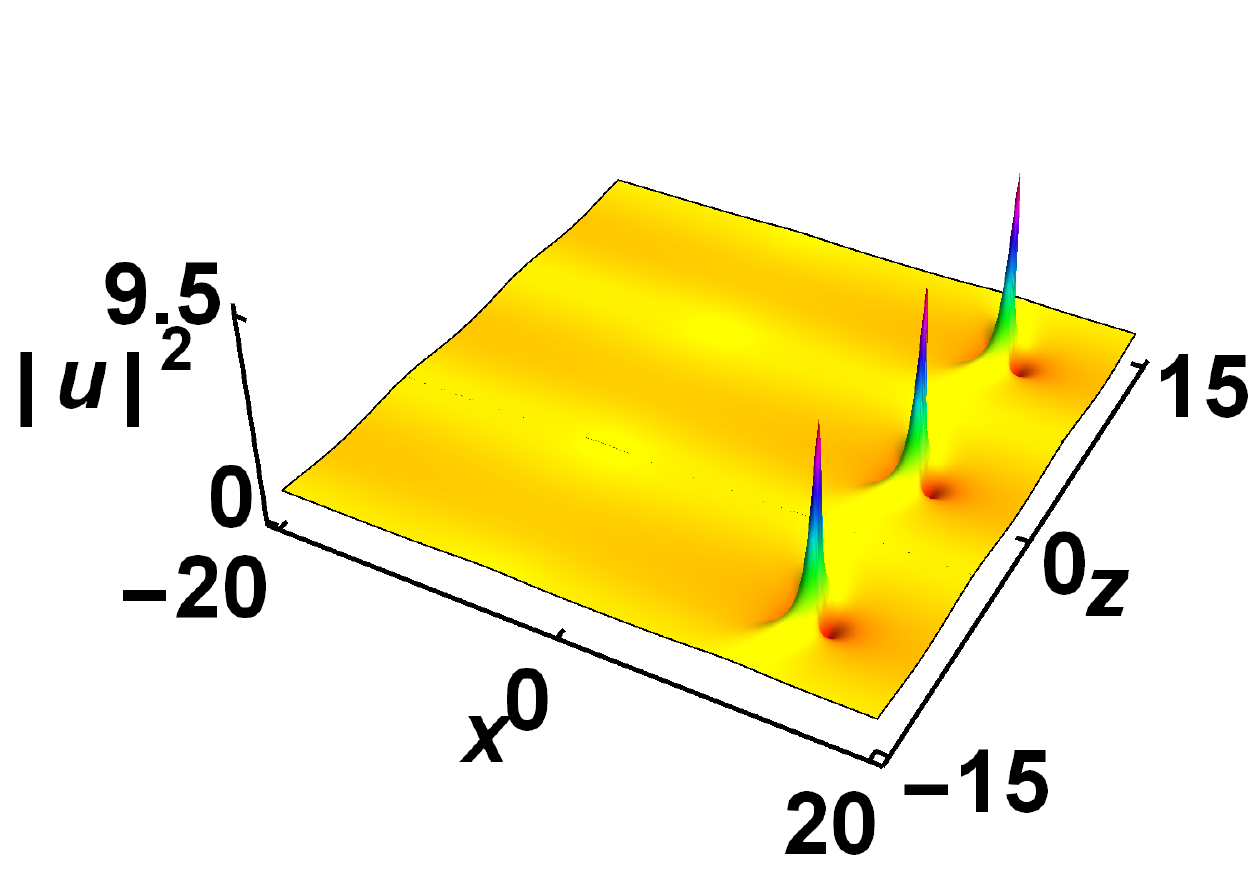}}
	\caption{Intensity distribution for controlled type [1] breathers with $\sigma=0.1$, $\omega=0.6$ and $a\neq0$, $b=0$. For: (a) $a=50$; (b) $a=100$; (c) $a=1500$.}
	\label{fig: fig4 }					
\end{figure}

To explain further, for specific modulating parameters, $\sigma=0.1$ and $\omega=0.6$, when the real part of the $s_1$ parameter, i.e. $a$ is increased under periodically modulated dispersion, the distance between the triplets of each unit is found to increase at first (see Figs. 4(a) and 4(b)). But after a certain $a$ value, the controlled three-peak type [1] breather transforms into a single-peak type [1] breather with a lower amplitude. The appearance of the single-peak type [1] breather is shown at a specific value $a=1500$ as presented in Fig. 4(c). This controlled single-peak type [1] breather appears similar to the controlled KM-like breather shown in Fig. 1(b), having its origin shifted to the positive $x$-axis. The reason behind the disappearance of the other two peaks may be attributed to the high $a$ value in comparison with the modulating period (2$\pi$/$\omega$) of the background. The higher $a$ value leads to a larger peak separation in each triplet unit of the controlled type [1] breather. In the presence of lower modulating period and larger peak separation (controlled by $a$), the other two peaks in each triplet unit stop breathing. When $a>1500$, the origin of the single-peak type [1] breather keeps shifting along the positive $x$-axis. 
\begin{figure}[ht]
		\centering
		\subfloat[]{\includegraphics[width=0.33\linewidth]{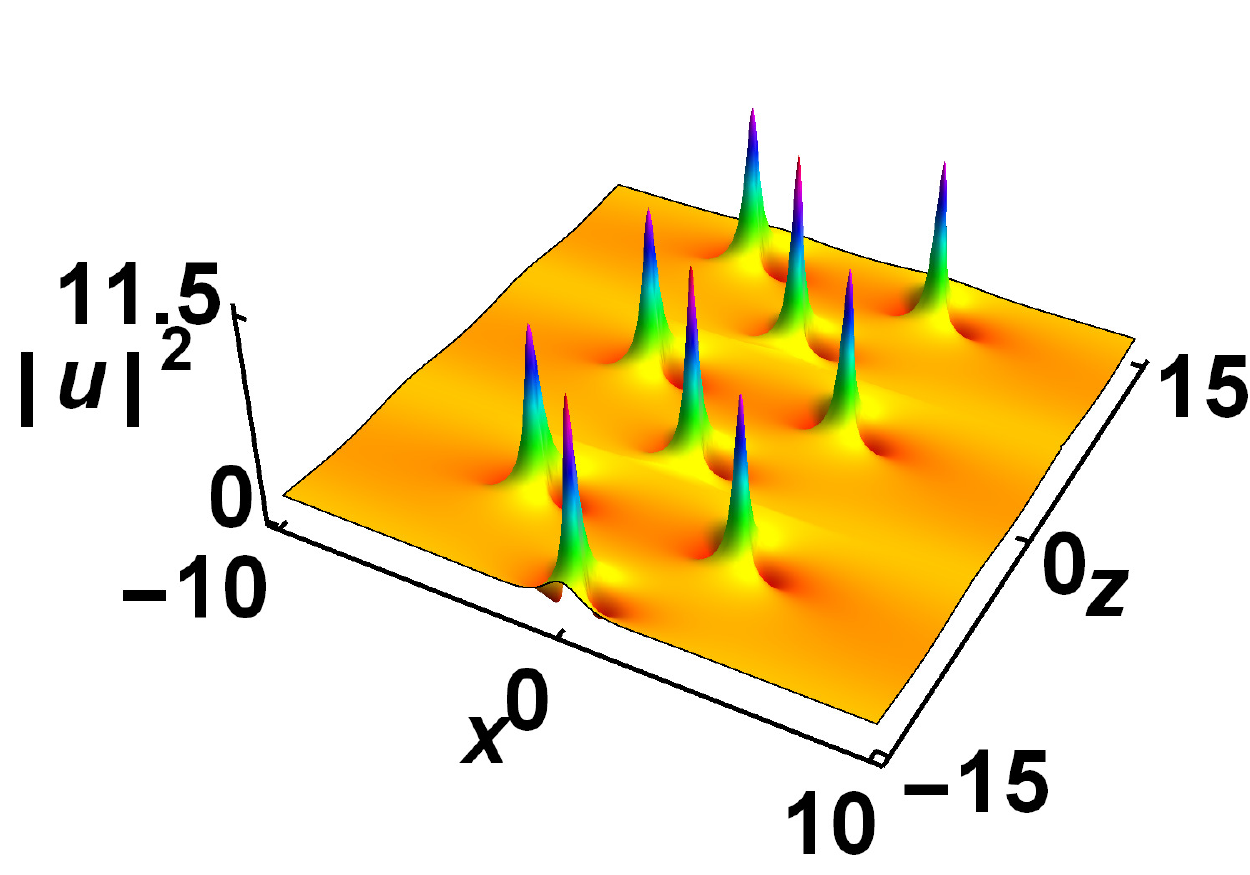}}
		\subfloat[]{\includegraphics[width=0.33\linewidth]{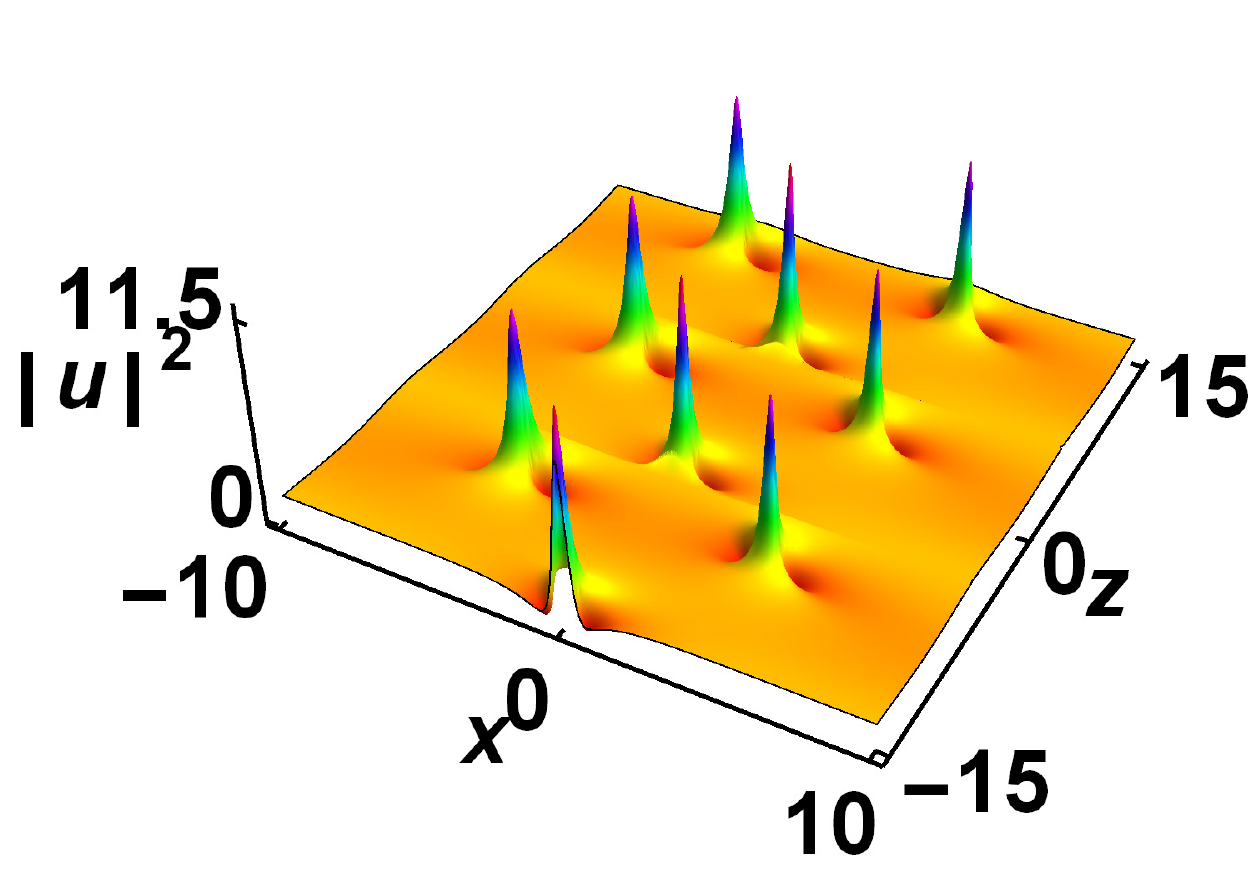}}
		\subfloat[]{\includegraphics[width=0.33\linewidth]{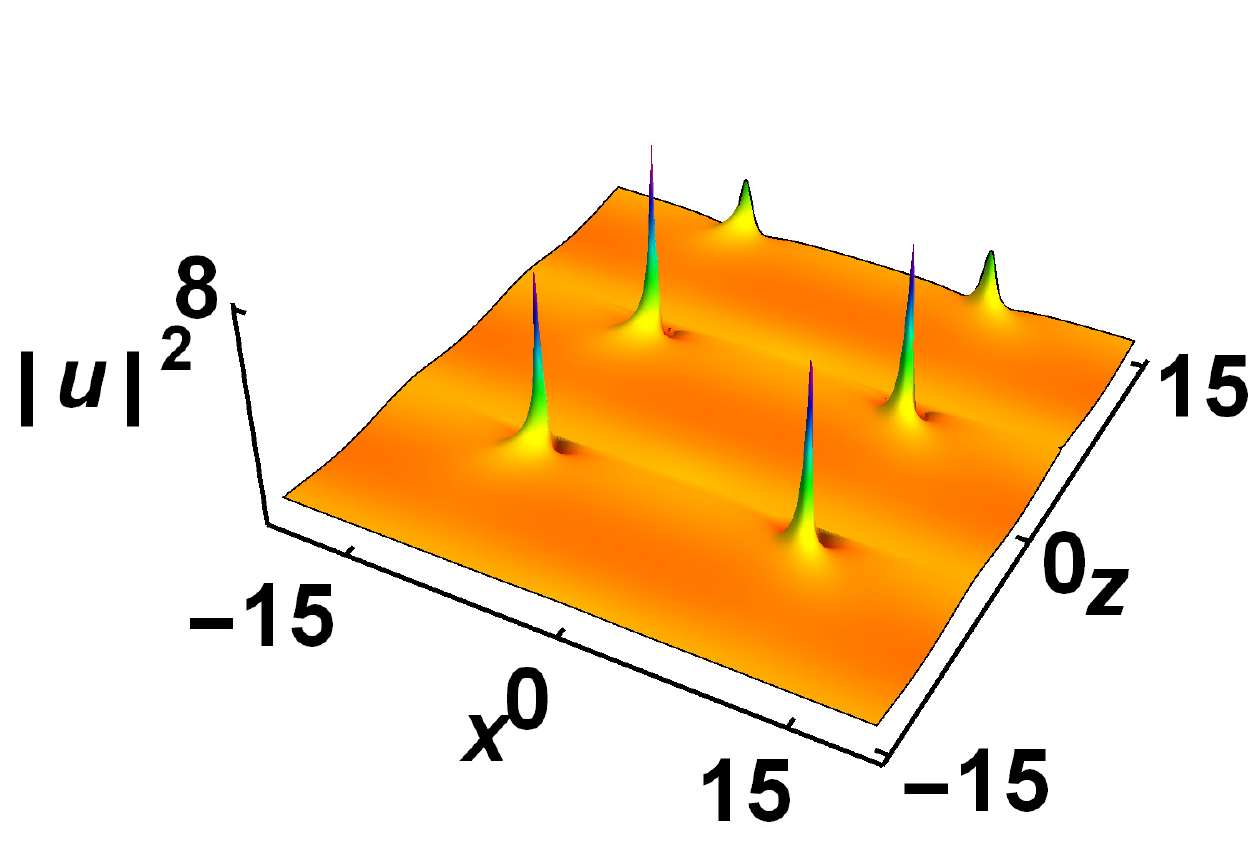}}
	\caption{Intensity distribution for controlled type [1] breathers with $\sigma=0.1$, $\omega=0.6$ and $a=0$, $b\neq0$.For: (a) $b=50$; (b) $b=100$; (c) $b=1500$.}
	\label{fig: fig5}				
\end{figure}
\begin{figure}[ht]
	\centering
	\includegraphics[width=0.45\linewidth]{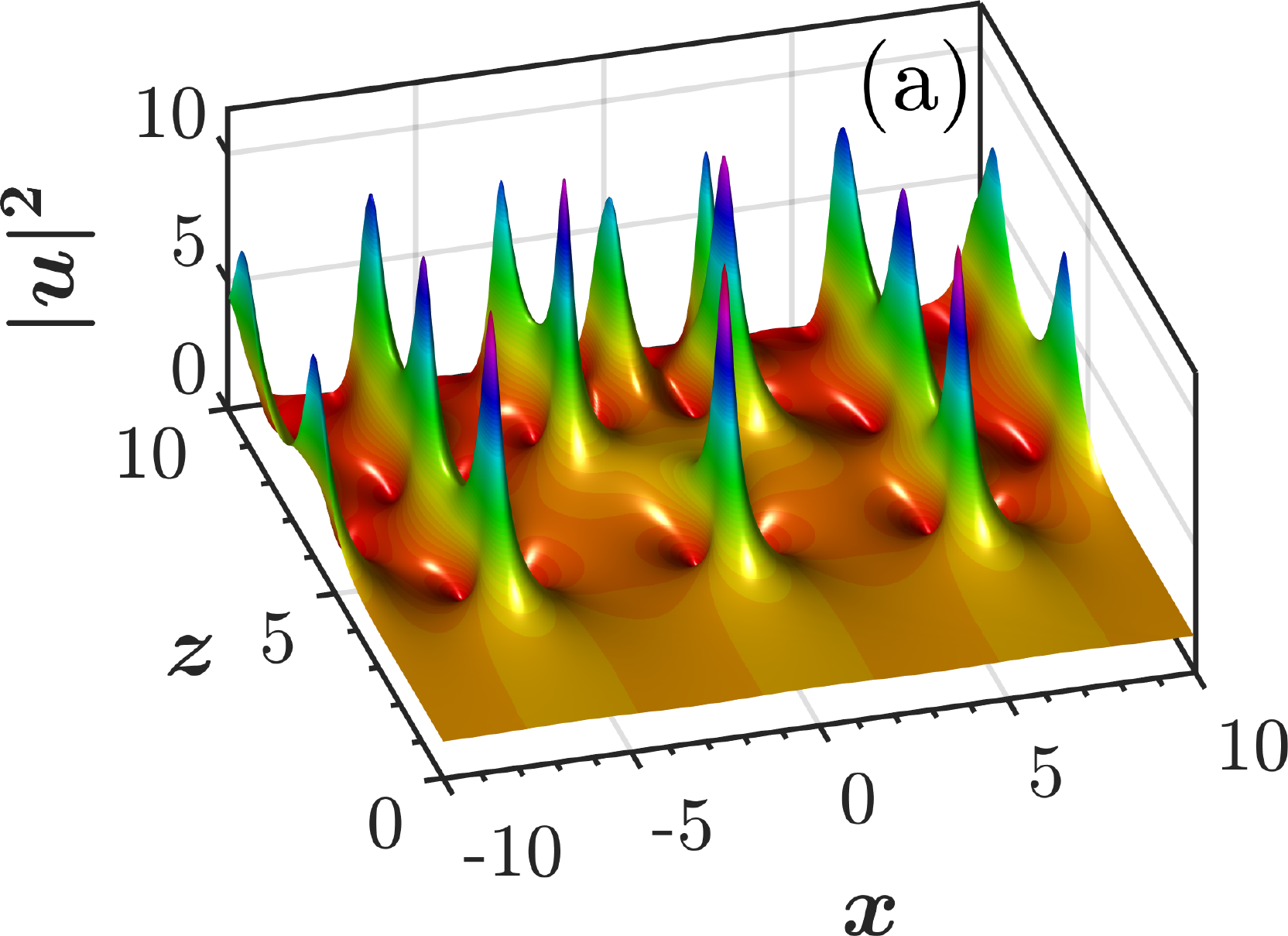}\includegraphics[width=0.45\linewidth]{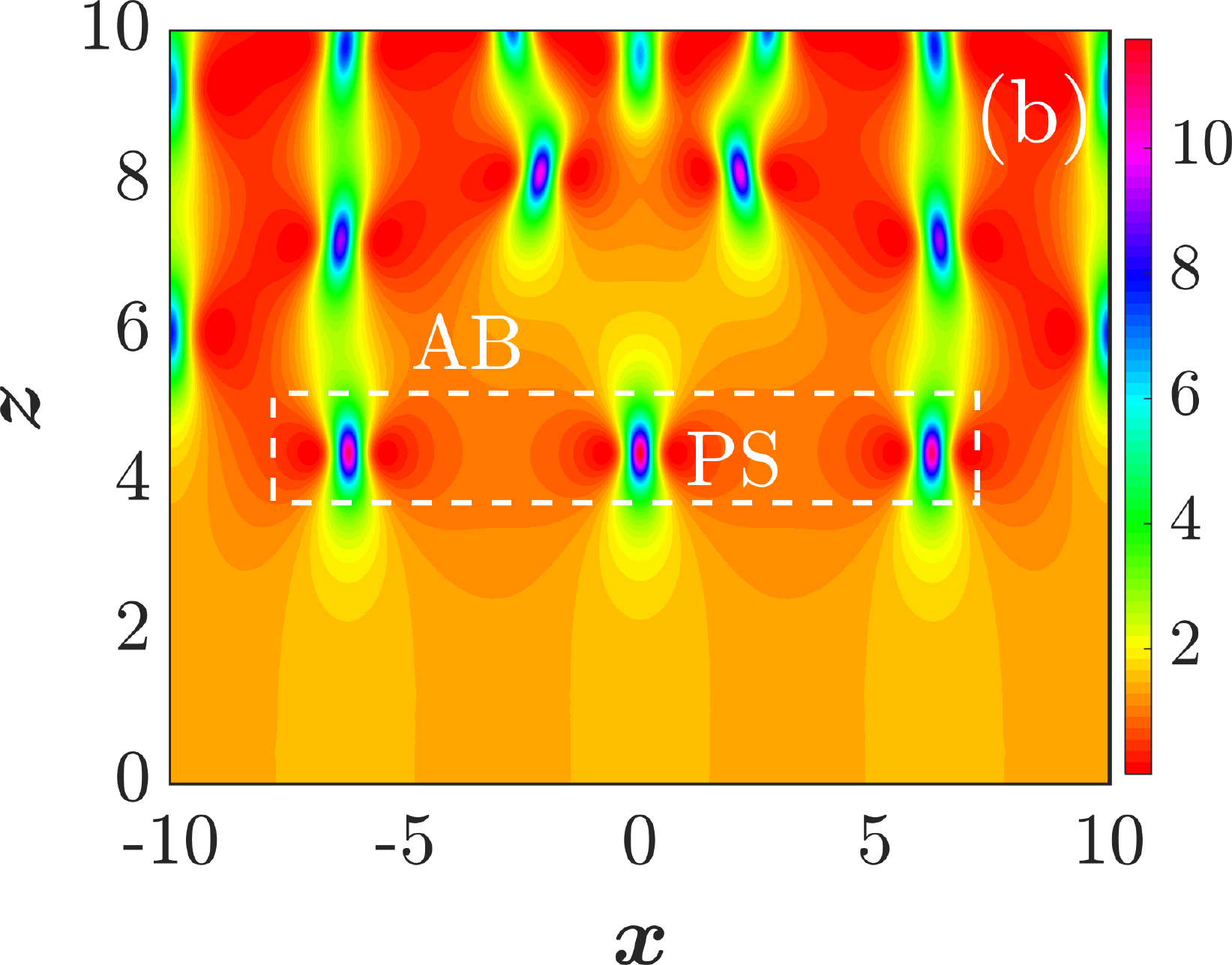}\\
	\includegraphics[width=0.4\linewidth]{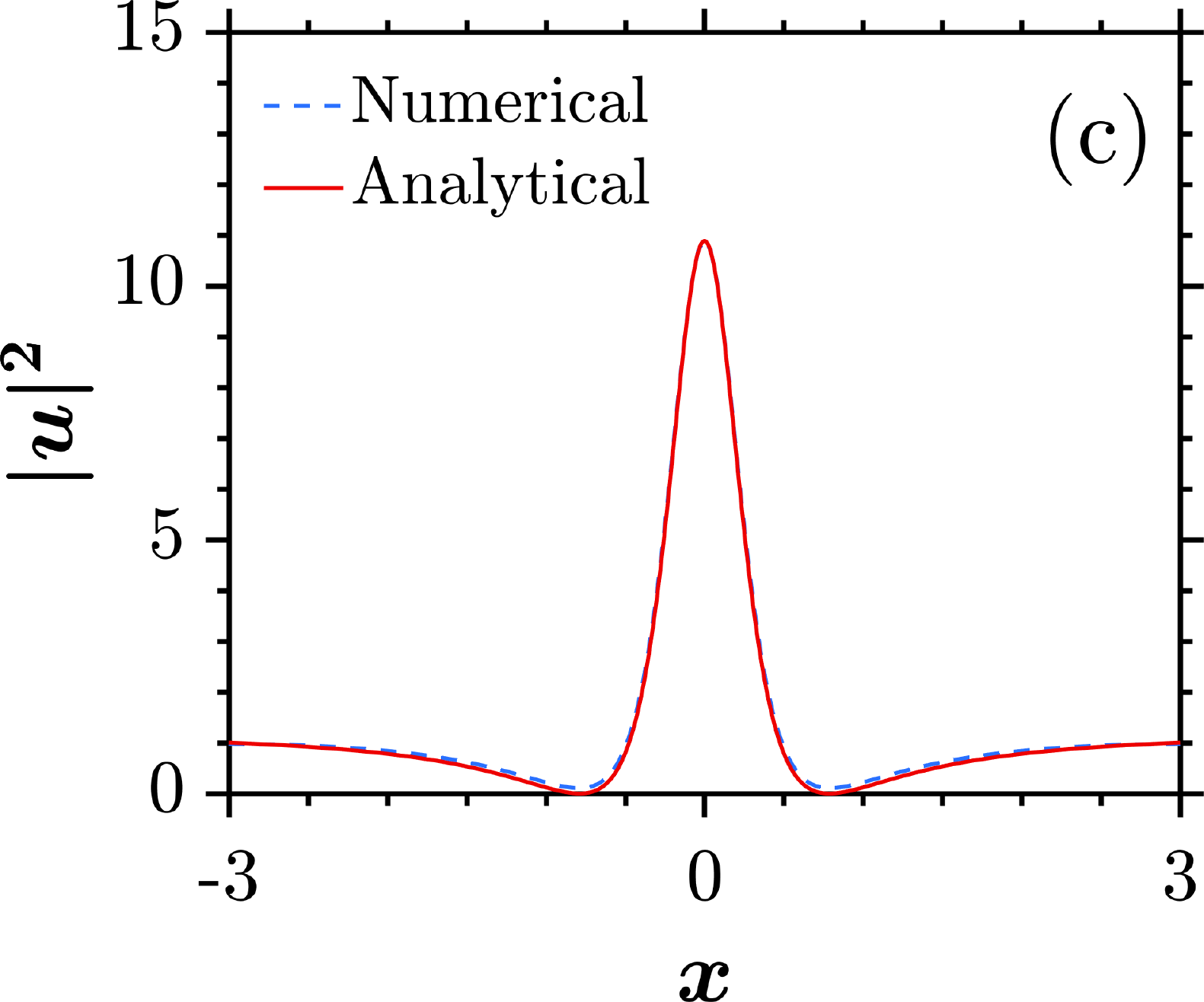}\includegraphics[width=0.48\linewidth]{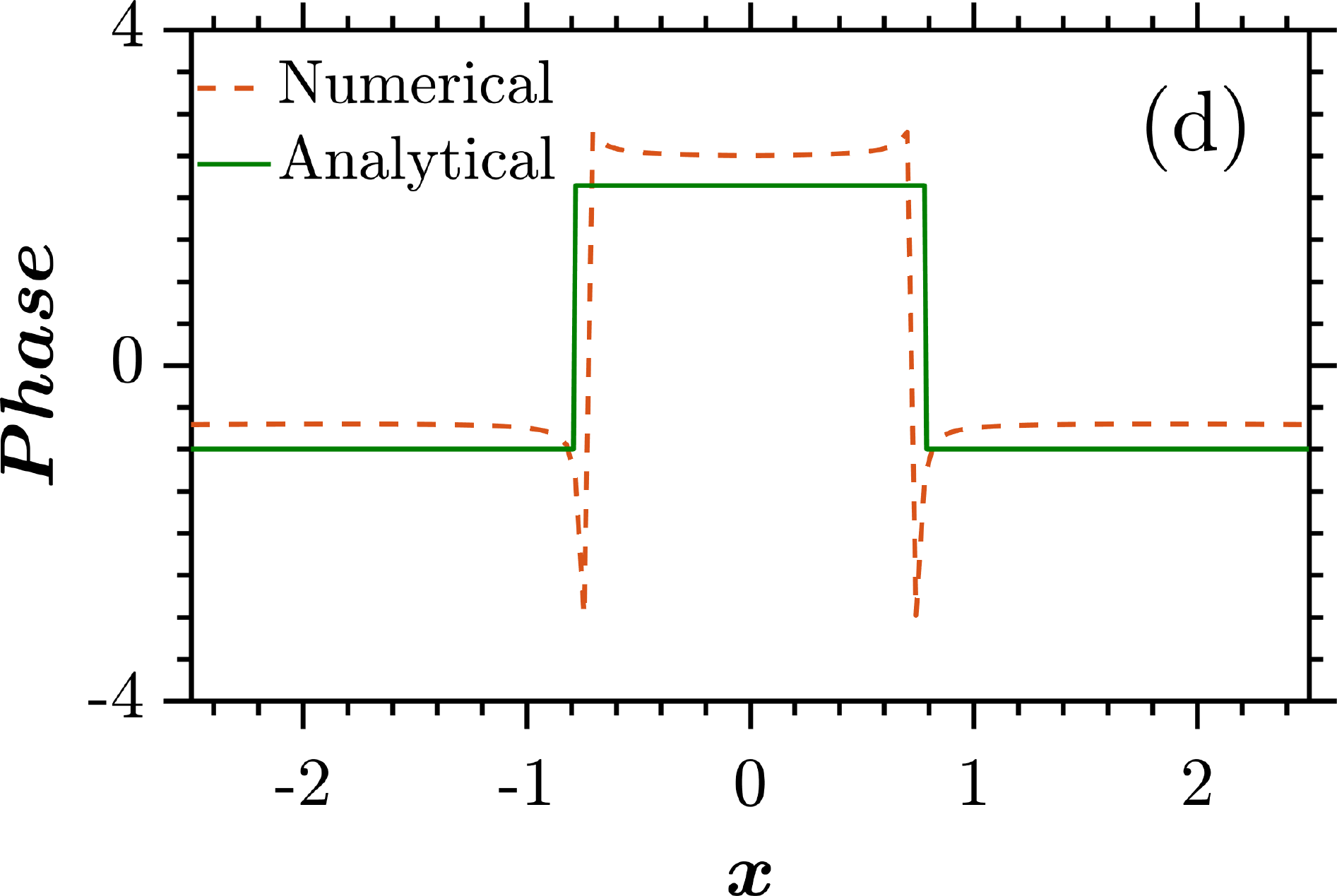}
	\caption{Numerical MI dynamics showing a sea of pulses for the case of standard NLS systems. The upper panels indicate the formation of AB-like states (indicated by a rectangle box drawn by dashed white line in (b)) induced by a small amplitude of perturbation ($\epsilon=0.01$) while the bottom panels (c) show a comparison of a single analytic PS and a localized state extracted from the density map at $z=4.5$ and (d) depict the corresponding phase portraits with the system parameters as $\beta_2(z)=\chi(z)=1$ and $\Omega=1$.}
\end{figure}

On the other hand, for the same modulating parameters $\sigma=0.1$ and $\omega=0.6$, when the imaginary part of $s_1$ parameter, i.e. $b$ is increased, the distance between the triplets within each unit is again found to increase as given in Figs. 5(a) and 5(b). But unlike the previous case, at a particular value $b=1500$, the transformation from three-peak type [1] breather into two-peak type [1] breather can be observed (see Fig. 5(c)). The breathing period of the third peak in each unit is less than the peak separation distance (controlled by $b$). In this case, another feature has also been observed that each unit gets shifted along the positive $z$-axis as shown in Fig. 5(c). When $b>1500$, peak separation in each unit increases and each unit shifts along the positive $z$-axis. It is well-known that the standard second-order rational solutions (type [0] and type [1]) do not possess two peaks, but by regulating the free parameters $b$ under periodic dispersion profile, we obtain two-peak type [1] breather solution. Thus, if the value of modulating period of finite background (controlled by spatial frequency in periodic dispersion profile) is less than the peak separation value (controlled by free parameters), one can obtain a new breather dynamics corresponding to the higher-order rational solutions.
\begin{figure}[ht]
	\centering
	\includegraphics[width=0.4\linewidth]{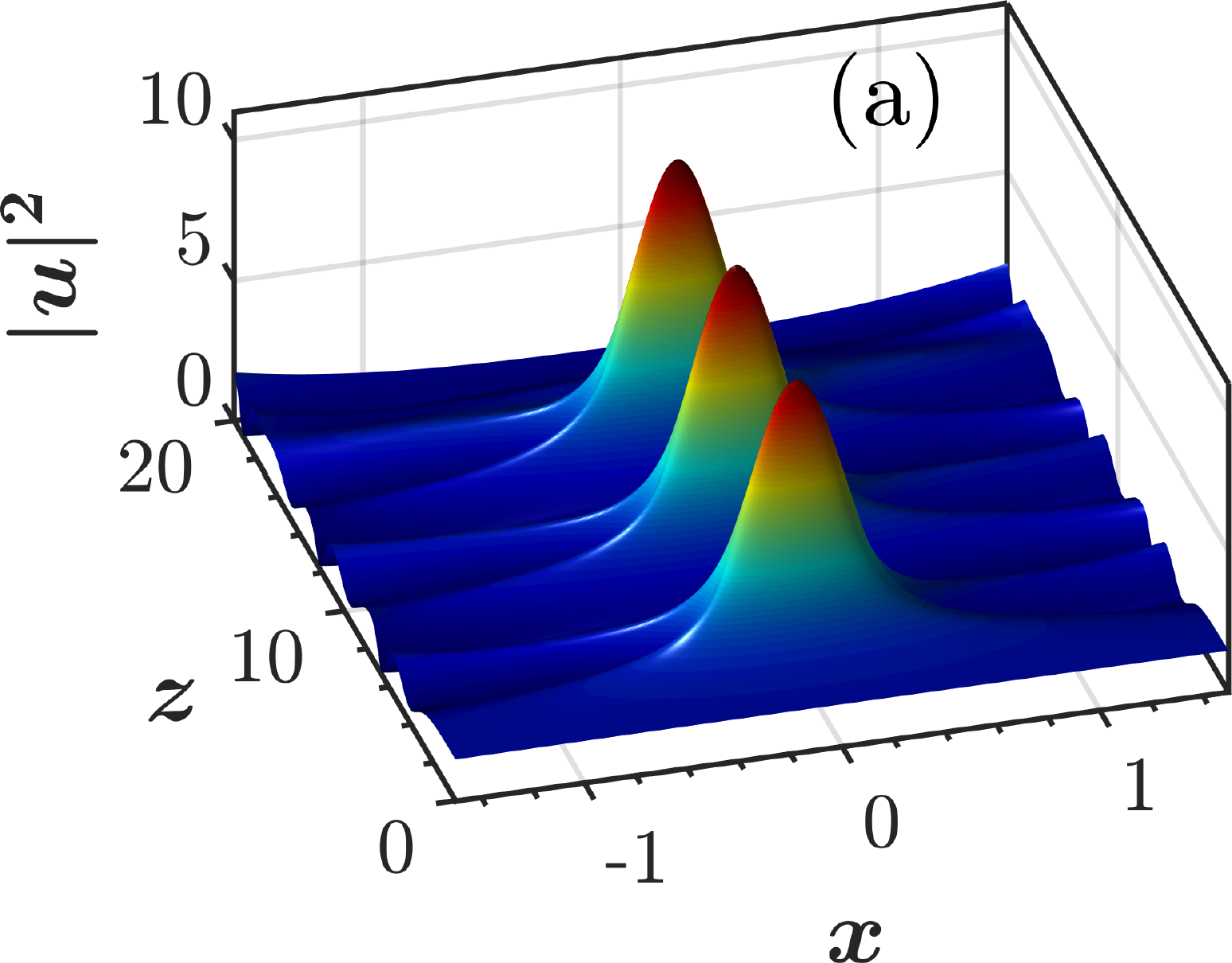}\hspace{0.6in}\includegraphics[width=0.4\linewidth]{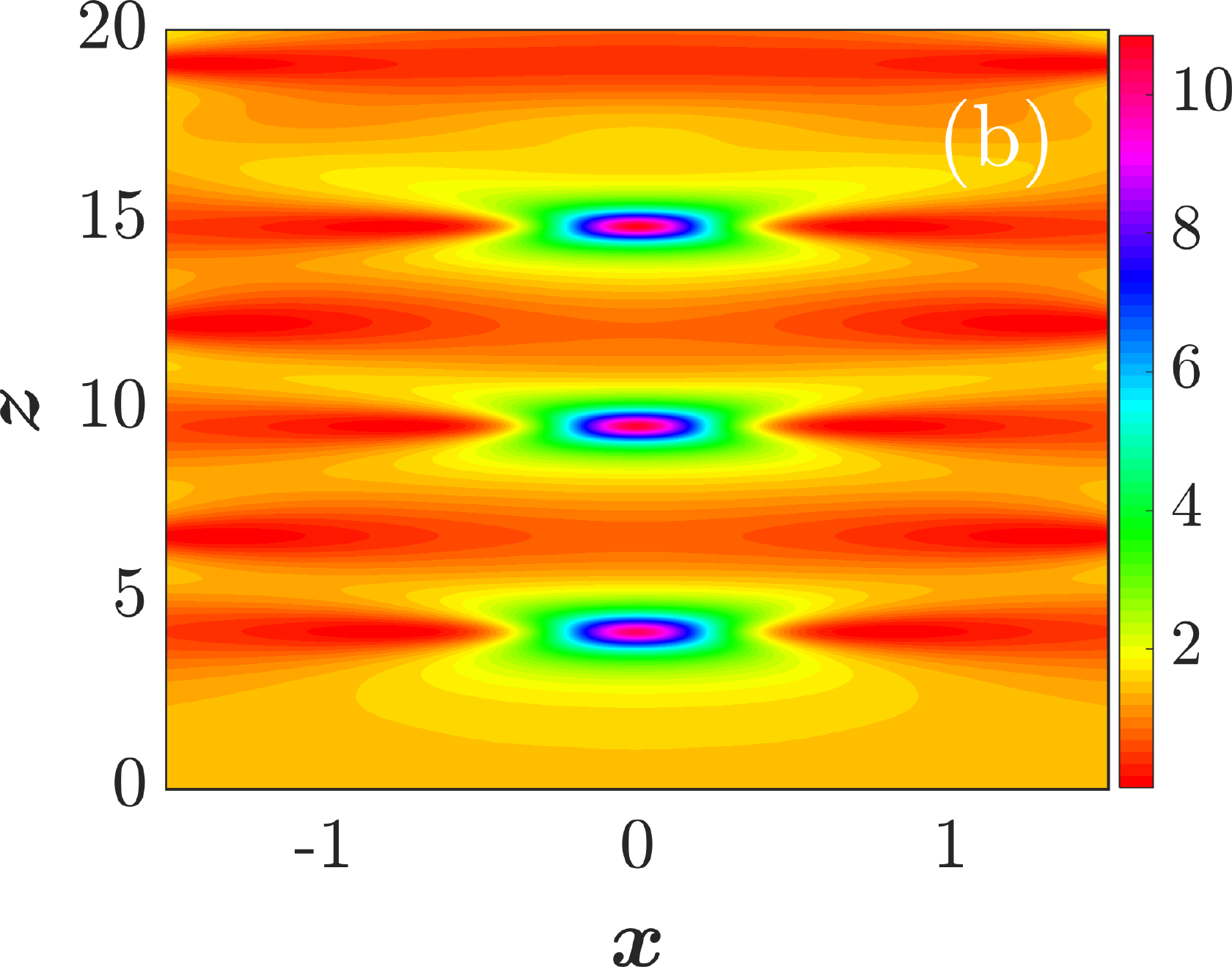}
	\includegraphics[width=0.4\linewidth]{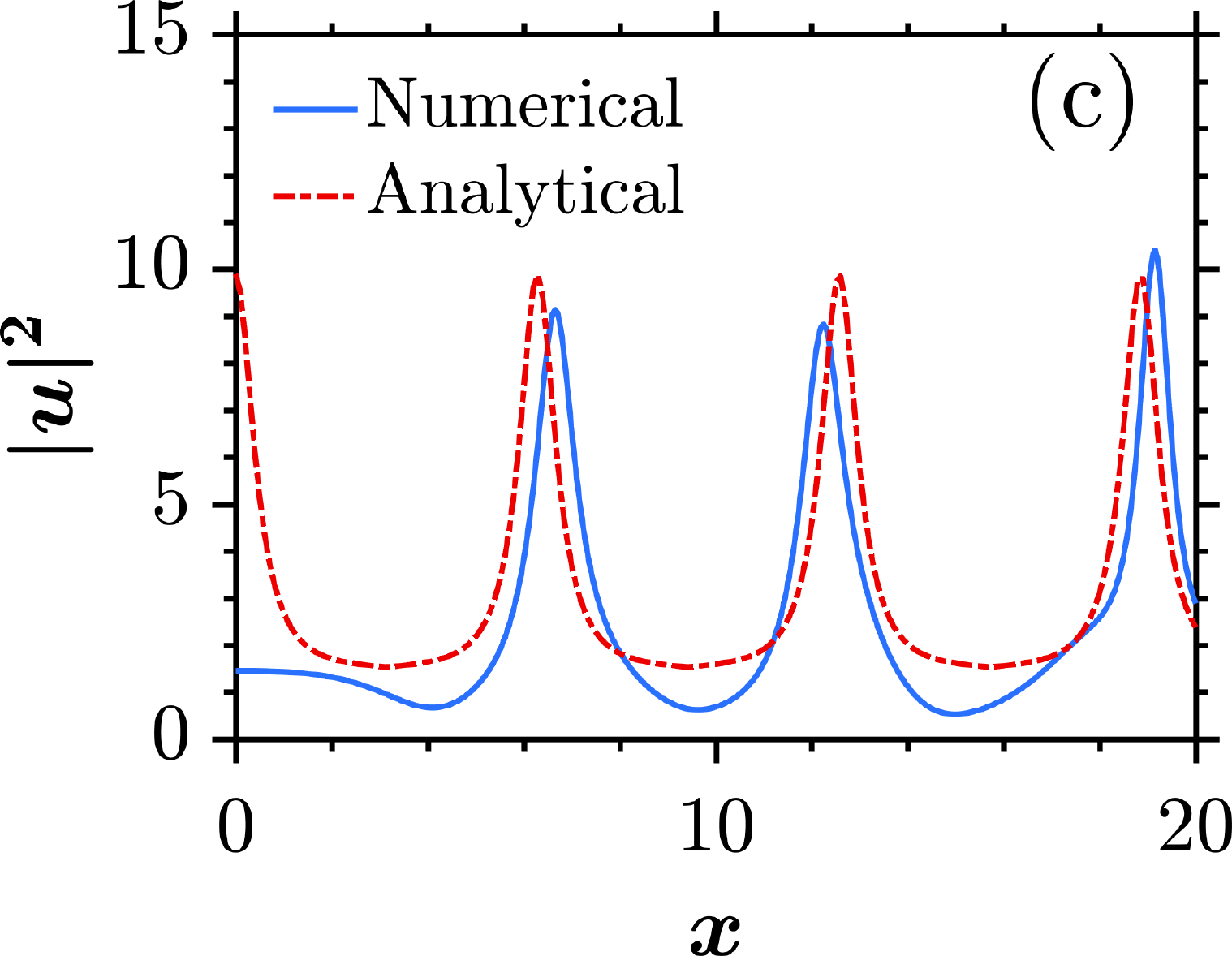}\hspace{0.2in}\includegraphics[width=0.4\linewidth]{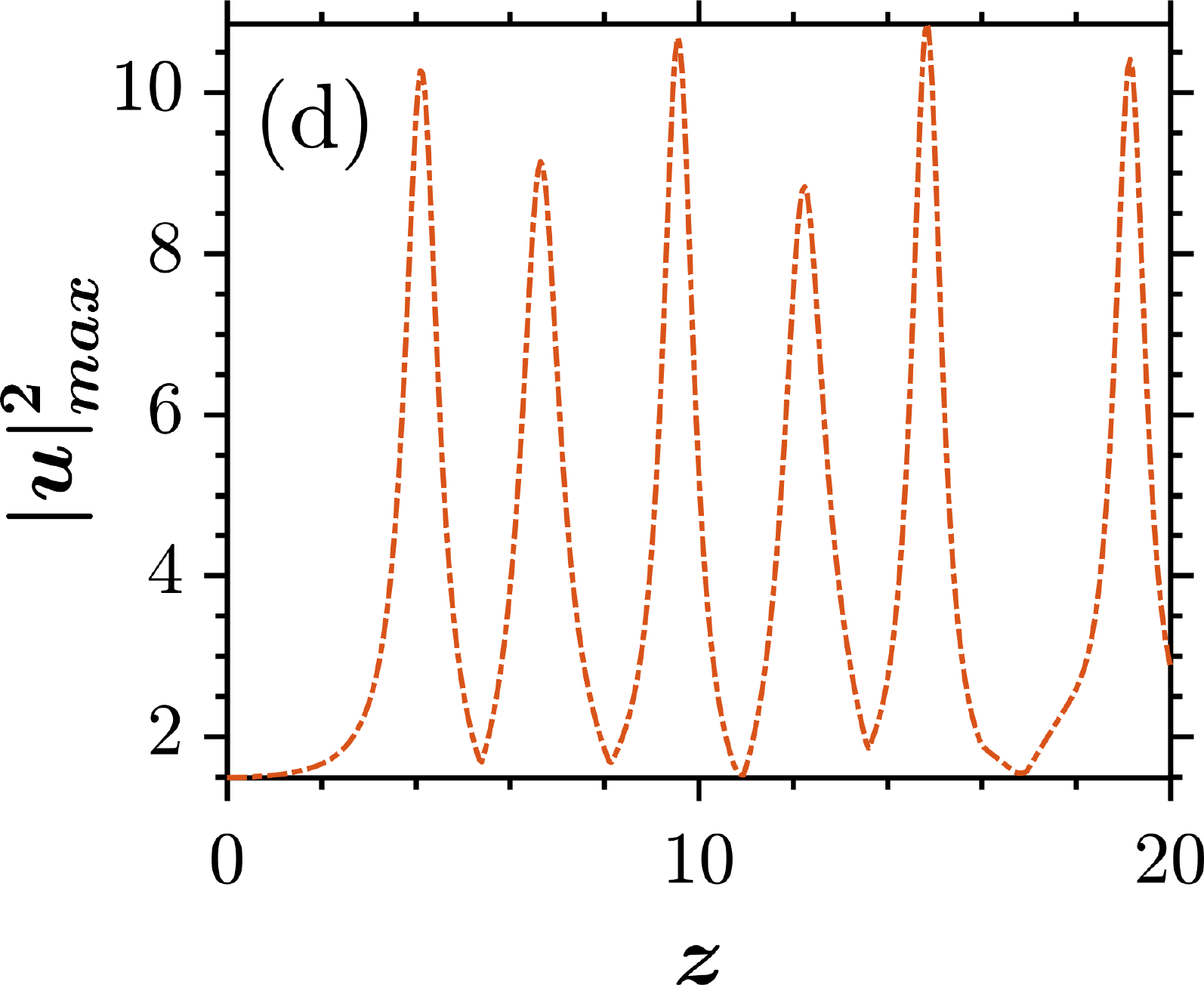}
	\caption{Dynamics of KM-like breathers by a train of PSs along the evolution co-ordinate is shown in upper panels. In bottom panels, (c) comparison of the localized state of PS extracted from the density map with the analytic KM breathers and (d) shows the growth rate against the small perturbation. Here, the parameters are the same as used in Fig. 1(b) with $\Omega=1$}.
\end{figure}
\section{\label{sec:level10}Modulational instability analysis: Direct numerical outcomes }

It is a well-established fact that modulational instability is a precursor for the formation of rogue waves as the latter is a phenomenon of extreme events triggered by the maximum unstable modes \cite{baronio2014vector}. Indeed, in realistic physical settings including optical fiber and sea waves the plane wave is perturbed by a quantum noise and subsequently the formation of extreme waves (rogue waves) is traced at a particular propagation distance. It is to be noted  that since the rogue waves are generated due to perturbative fields, they are highly unstable which in turn results to an instantaneous life time to the former. Hence considering the analytical solutions as the seed solutions in the numerical evolution are unrealistic ones. One of the standard ways to simulate these analytical solutions is to rely on the noise induced MI where the the plane wave experiences an instability due to the perturbation \cite{ablowitz2001long}. In this process, the spectral sidebands of unstable modes undergo an exponential growth and eventually lead to a more complex dynamics which could be mapped to the well-known four-wave mixing process. During this complex dynamics, there occurs a cyclic energy exchange between multiple spectral modes due to an inelastic collision \cite{agrawal2013nonlinear}. Besides, it has been suggested that the occurrence of MI is closely related to the phenomenon of Fermi–Pasta–Ulam recurrence in optics since the MI, in the long run, results in a train of ultra-short pulses with a same repetition rate in its chaotic MI map consisting of multiple high- amplitude waves which can show a signature of extreme events \cite{van2001}. To observe the realtime rogue waves investigated in the proposed system of periodically varying group dispersion, we solve Eq. (1) by the standard pseudo-spectral method in \textit{Python } programming language with appropriate boundary conditions including 1024 Fourier points \cite{govindarajan2017}. To this end, we assume the following plane wave solution with an infinitesimal parameter $\epsilon$, which is to be accountable for the perturbation. 
\begin{equation}
u(0,x)=u_0(1+\epsilon \cos(\Omega x)).
\end{equation}
where $\Omega$ is the (temporal) frequency of the applied field. Here the values of CW amplitude and the seed parameter are, respectively, chosen to be $u_0=1.21$ and $\epsilon=0.01$ unless otherwise stated. Prior to analyzing the generation of rogue waves in the inhomogeneous system, we first produce the results for the constant NLSE by considering the periodic GVD parameter as  $\beta_2(z)=1$ by keeping the nonlinear parameter  as $\chi(z)=1$. As shown in Fig. 6(a), Akhmediev-like breathers (ABs) are formed  after a finite distance ($z>3$) (see Fig. 6(b)) before which the non-zero CW background does not experience any instability. Note that these ABs are generated by an array of Peregrine solitons (refer to the pulse marked with a white rectangle box in Fig. 6(b) and the one dimensional plot drawn in Fig. 6(c)) confined in temporal axis ($x$) and these results well agree with the earlier analytical predictions shown in Fig. 1(a). While making this conclusion, it must be kept in mind that both the amplitude and phase of the nonlinear structure  are needed to arrive at a reasonable inference in order to prove the typical signatures of PSs \cite{randoux2016inverse}. In Fig. 6(d), we have shown the phase portraits of both the analytical solution shown in Fig. 1(a) and the numerically extracted PSs (for instance, see Fig. 6(b)) at the propagation distance $z=4.5$. It is obvious that the outcomes driven by the modulational instability of plane wave induced by the perturbation agree well with the measured intensity and phase of the nonlinear wave packets as were done in Refs. \cite{akhmediev2009extreme, dudley2014}

We now investigate the dynamics of Peregrine solitons in the case of inhomogeneous systems with the periodic profile ($\beta_2(z)=1+\sigma \cos(\omega z)$). By retaining the same parameters used in Fig. 1(b) with $\sigma=0.1$ and $\Omega=1$, a series of numerical experiments have been carried out by varying the temporal domain ($x$) to capture the formation of KM- like breathers and the corresponding ramifications shown in Figs. 7(a) and 7(b)  corroborate well with the analytical calculations depicted in Fig. 1(b). In support of this validation, a comparison of the two dimensional intensity plot of Fig. 1 (b)  with the evolution of the plane wave along the propagation direction is shown in  Fig. 7(c). It is evident from Fig. 7(c) that the obtained numerical KM- like breathers are a clear manifestation of periodic management of dispersion parameter. In addition, the growth rate against the evolution co-ordinate is given in Fig. 7(d), which further supports the presented numerical findings adequately. Although the first-order rational solutions can be traced after a number of numerical experiments pertaining to the modulational instability caused by the small perturbation, it is highly cumbersome to predict the existence of higher-order Peregrine solitons \cite{dudley2014}. Nevertheless, we have presented the dynamics of third-order rational solutions in Fig.  8 with the system parameters chosen as $\sigma=0.125$ and $\omega=1$. It is very clear to observe the formation of claw-like structure (refer to the PS indicated by a red arrow line in Fig. 8(a)) with the triplets of Peregrine solitons. We affirm here that, to the best of our knowledge, these are the first ever numerical proofs for the existence of Peregrine solitons and KM breathers done in the inhomogeneous systems through the direct simulations induced by the infinitesimal perturbations. Also, it is quite remarkable to note that such perturbation induced nonlinear structures have been mapped well with the analytical solutions through the MI simulations instead of simulating the former along the evolution co-ordinate by considering it as the seed solution even in this inhomogeneous system. It is worthwhile to mention that such PSs and their different variants including ABs and KM-like breathers found both analytically and numerically in the present system can be characterized and distinguished by some other import tools such as numerical inverse scattering transform (IST) and finite gap theory \cite{ablowitz2001long,ablowitz1996computational,grinevich2018finite,randoux2016}. In particular, IST spectra obtained from the eigenvalues of the considered systems stand out to be an accurate tool in characterizing extreme waves (rogue waves) and their mechanisms. Our analytical and numerical findings are also expected to rely on this method and those ramifications shall be published elsewhere. Since the periodic GVD can easily be fabricated by the state-of-the-art-technology available in this modern era and these novel solutions have already been observed through the numerical experiments,   we do hope that the present study will stimulate experiments in these directions, which may pave a new avenue on rogue waves in the lightwave communication systems.
\begin{figure}[ht]
	\centering
	\includegraphics[width=0.4\linewidth]{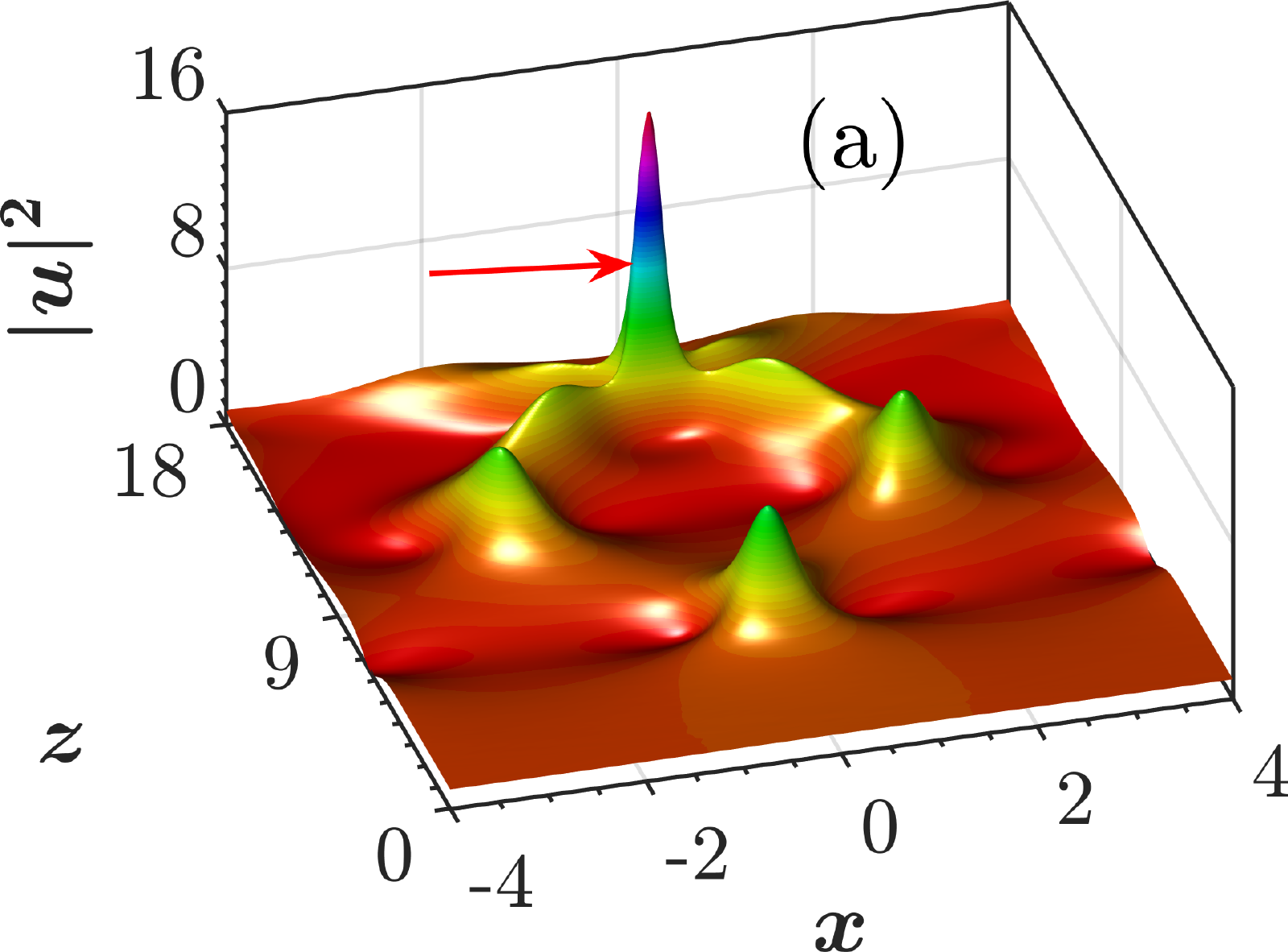}\hspace{0.4in}\includegraphics[width=0.38\linewidth]{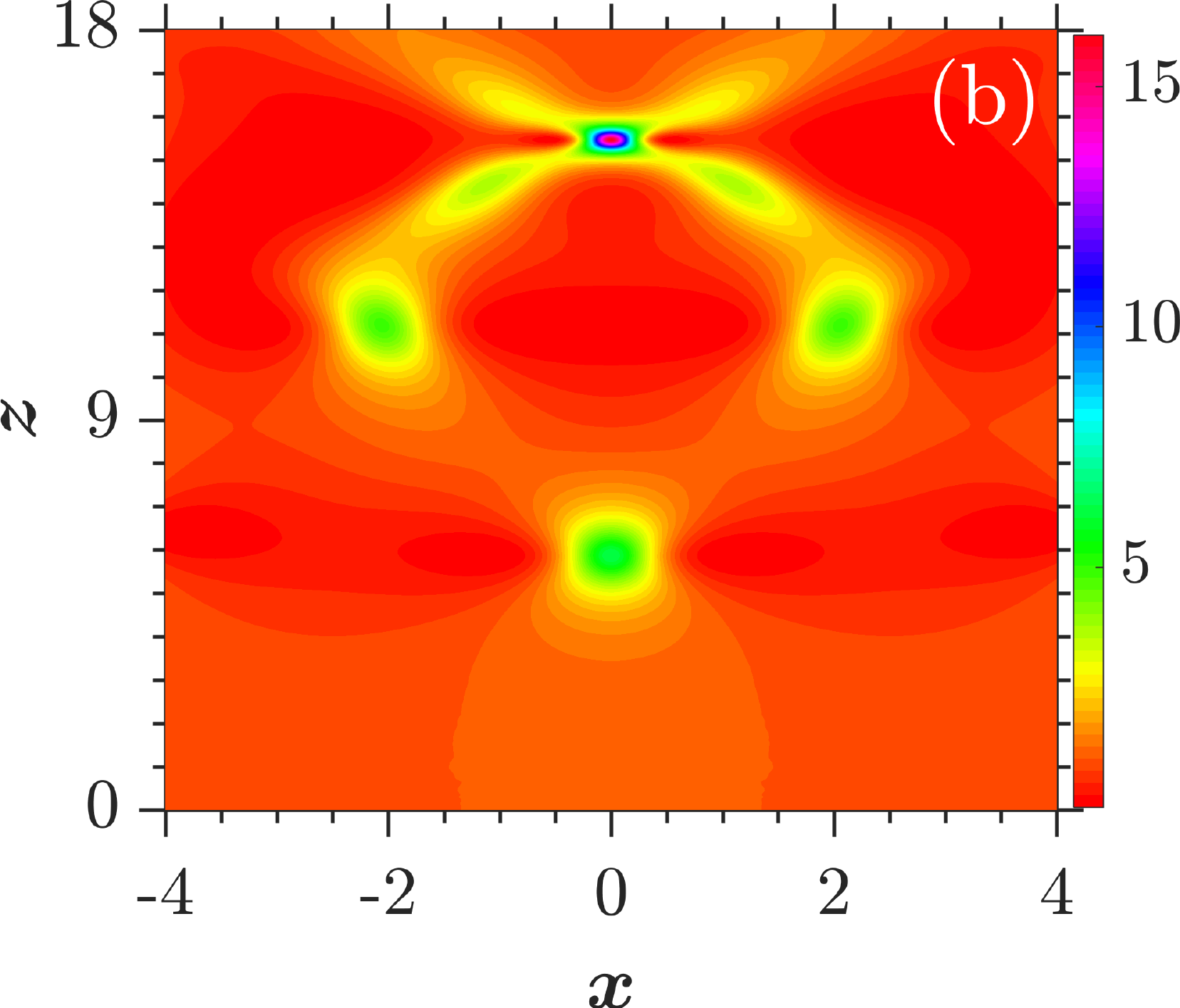}
	\caption{Evolution of third-order rational waves with type [1,1] induced by the amplitude of perturbation  $\epsilon=0.01$. The other parameters are $\sigma=0.125$, $\omega=1$ and $\chi=\Omega=1$.}
\end{figure}

\section{\label{sec:level11}Conclusion}
In summary, the evolution of first three orders of rational soliton solutions has been investigated under the presence of a longitudinally varying periodic dispersion via appropriate similarity transformation in the context of optical fiber. It has been shown that it is possible to transform rational soliton solution into KM-like breather with judicious choice of dispersion coefficient parameters. By modulating the amplitude part and the spatial frequency of the periodic dispersion profile, one can realize novel rational solution dynamics. Our study also shows how for periodic dispersion, the second-order rational solution dynamics change depending on free parameters corresponding to the complex $s_1$ parameter. For specific modulating parameters, different peak-dynamics have been observed by regulating the real and imaginary parts of the $s_1$ parameter. Direct numerical simulations have also been executed to corroborate the analytical predictions through the modulational instability analysis by perturbing the applied field with a small perturbation in the form of white noise. Since in reality modulating dispersion profile is much easier, as evident from numerous experimental studies, than modulating nonlinearity or any other inhomogeneous profiles, we believe our study on the dynamics of rational solutions  can be helpful to understand the dynamics of different types of rogue waves in realtime experiments. 

\section*{Declaration of competing interest}
The authors declare that they have no known competing financial interests or personal relationships that could have appeared to influence the work reported in this paper.

\section*{Acknowledgement}
 
 D.K.M. thanks MHRD, Government of India for financial support through a fellowship. A.K.S. acknowledges financial support from  Science and Engineering Research Board (SERB), Government of India under MATRICS scheme (Grant No. MTR/2019/000945). A.G. and M.L. acknowledge the support of Science and Engineering Research Board (DST-SERB) for providing a Distinguished Fellowship project to  M.L. (Grant No. SB/DF/O4/2017) in which AG was a Visiting Scientist. A.G. is now supported by University Grants Commission (UGC), Government of India, through a Dr. D. S. Kothari Postdoctoral Fellowship (Grant No. F.4-2/2006 (BSR)/PH/19-20/0025).


\begin{thebibliography}{76}
	\bibitem{solli2007optical}
	D.~Solli, C.~Ropers, P.~Koonath, B.~Jalali, Optical rogue waves, Nature
	450~(7172) (2007) 1054.
	
	\bibitem{kibler2010peregrine}
	B.~Kibler, J.~Fatome, C.~Finot, G.~Millot, F.~Dias, G.~Genty, N.~Akhmediev,
	J.~M. Dudley, The Peregrine soliton in nonlinear fibre optics, Nature Physics
	6~(10) (2010) 790--795.
	
	\bibitem{dudley2015rogue}
	J.~M. Dudley, M.~Erkintalo, G.~Genty, Rogue waves of light, Optics and
	Photonics News 26~(11) (2015) 34--41.
	
	\bibitem{bludov2013instabilities}
	Y.~V. Bludov, R.~Driben, V.~V. Konotop, B.~Malomed, Instabilities, solitons and
	rogue waves in $\mathcal{PT}$-coupled nonlinear waveguides, Journal of Optics 15~(6)
	(2013) 064010.
	
	\bibitem{onorato2013rogue}
	M.~Onorato, S.~Residori, U.~Bortolozzo, A.~Montina, F.~Arecchi, Rogue waves and
	their generating mechanisms in different physical contexts, Physics Reports
	528~(2) (2013) 47--89.
	
	\bibitem{kibler2012observation}
	B.~Kibler, J.~Fatome, C.~Finot, G.~Millot, G.~Genty, B.~Wetzel, N.~Akhmediev,
	F.~Dias, J.~M. Dudley, Observation of kuznetsov-ma soliton dynamics in
	optical fibre, Scientific Reports 2 (2012) 463.
	
	\bibitem{baronio2020resonant}
	F.~Baronio, S.~Chen, S.~Trillo, Resonant radiation from peregrine solitons,
	Optics Letters 45~(2) (2020) 427--430.
	
	\bibitem{xu2019breather}
	G.~Xu, A.~Gelash, A.~Chabchoub, V.~Zakharov, B.~Kibler, Breather wave
	molecules, Phys. Rev. Lett. 122~(8) (2019) 084101.
	
	\bibitem{kraych2019nonlinear}
	A.~E. Kraych, P.~Suret, G.~El, S.~Randoux, Nonlinear evolution of the locally
	induced modulational instability in fiber optics, Phys. Rev. Lett. 122~(5)
	(2019) 054101.
	
	\bibitem{scott1982dynamics}
	A.~C. Scott, Dynamics of Davydov solitons, Physical Review A 26~(1) (1982) 578.
	
	\bibitem{scott1984magma}
	D.~R. Scott, D.~J. Stevenson, Magma solitons, Geophysical Research Letters
	11~(11) (1984) 1161--1164.
	
	\bibitem{stenflo2010rogue}
	L.~Stenflo, M.~Marklund, Rogue waves in the atmosphere, Journal of Plasma
	Physics 76~(3-4) (2010) 293--295.
	
	\bibitem{zakharov1968stability}
	V.~E. Zakharov, Stability of periodic waves of finite amplitude on the surface
	of a deep fluid, Journal of Applied Mechanics and Technical Physics 9~(2)
	(1968) 190--194.
	
	\bibitem{lake1977nonlinear}
	B.~M. Lake, H.~C. Yuen, H.~Rungaldier, W.~E. Ferguson, Nonlinear deep-water
	waves: theory and experiment. part 2. evolution of a continuous wave train,
	Journal of Fluid Mechanics 83~(1) (1977) 49--74.
	
	\bibitem{haus1996solitons}
	H.~A. Haus, W.~S. Wong, Solitons in optical communications, Reviews of Modern
	Physics 68~(2) (1996) 423.
	
	\bibitem{kivshar2003optical}
	Y.~S. Kivshar, G.~Agrawal, Optical solitons: from fibers to photonic crystals,
	Academic press, 2003.
	
	\bibitem{dalfovo1999pitaevskii}
	F.~Dalfovo, S.~Giorgini, P.~Lev, Pitaevskii, and s. stringari, Rev. Mod. Phys
	71 (1999) 463.
	
	\bibitem{bludov2009matter}
	Y.~V. Bludov, V.~Konotop, N.~Akhmediev, Matter rogue waves, Physical Review A
	80~(3) (2009) 033610.
	
	\bibitem{wen2011matter}
	L.~Wen, L.~Li, Z.-D. Li, S.-W. Song, X.-F. Zhang, W.~Liu, Matter rogue wave in
	Bose-Einstein condensates with attractive atomic interaction, The European
	Physical Journal D 64~(2-3) (2011) 473--478.
	
	\bibitem{rajendran2010}
	S.~Rajendran, P.~Muruganandam, M.~Lakshmanan, Bright and dark solitons in a
	quasi-1d Bose--Einstein condensates modelled by 1d Gross--Pitaevskii equation
	with time-dependent parameters, Physica D: Nonlinear Phenomena 239~(7) (2010)
	366--386.
	
	\bibitem{rajendran2011}
	S.~Rajendran, M.~Lakshmanan, P.~Muruganandam, Matter wave switching in
	bose--einstein condensates via intensity redistribution soliton interactions,
	Journal of Mathematical Physics 52~(2) (2011) 023515.
	
	\bibitem{shukla2010nonlinear}
	P.~K. Shukla, B.~Eliasson, Nonlinear aspects of quantum plasma physics,
	Physics-Uspekhi 53~(1) (2010) 51.
	
	\bibitem{bailung2011observation}
	H.~Bailung, S.~Sharma, Y.~Nakamura, Observation of peregrine solitons in a
	multicomponent plasma with negative ions, Physical Review Letters 107~(25)
	(2011) 255005.
	
	\bibitem{shukla2012alfvenic}
	P.~Shukla, W.~Moslem, Alfv{\'e}nic rogue waves, Physics Letters A 376~(12-13)
	(2012) 1125--1128.
	
	\bibitem{gan2016solitons}
	J.-H. Gan, H.~Xiong, L.-G. Si, X.-Y. L{\"u}, Y.~Wu, Solitons in optomechanical
	arrays, Optics Letters 41~(12) (2016) 2676--2679.
	
	\bibitem{akhmediev1997solitons}
	N.~N. Akhmediev, A.~Ankiewicz, Solitons: nonlinear pulses and beams, Chapman \&
	Hall, 1997.
	
	\bibitem{shabat1972exact}
	A.~Shabat, V.~Zakharov, Exact theory of two-dimensional self-focusing and
	one-dimensional self-modulation of waves in nonlinear media, Soviet Physics
	JETP 34~(1) (1972) 62.
	
	\bibitem{ma1979perturbed}
	Y.-C. Ma, The perturbed plane-wave solutions of the cubic Schr{\"o}dinger
	equation, Studies in Applied Mathematics 60~(1) (1979) 43--58.
	
	\bibitem{akhmediev1986modulation}
	N.~Akhmediev, V.~Korneev, Modulation instability and periodic solutions of the
	nonlinear schr{\"o}dinger equation, Theoretical and Mathematical Physics
	69~(2) (1986) 1089--1093.
	
	\bibitem{peregrine1983water}
	D.~Peregrine, Water waves, nonlinear schr{\"o}dinger equations and their
	solutions, The ANZIAM Journal 25~(1) (1983) 16--43.
	
	\bibitem{cuevas2018stabilization}
	J.~Cuevas-Maraver, B.~A. Malomed, P.~G. Kevrekidis, D.~J. Frantzeskakis,
	Stabilization of the Peregrine soliton and Kuznetsov--Ma breathers by means
	of nonlinearity and dispersion management, Physics Letters A 382~(14) (2018)
	968--972.
	
	\bibitem{akhmediev1985generation}
	N.~Akhmediev, V.~Eleonskii, N.~Kulagin, Generation of periodic trains of
	picosecond pulses in an optical fiber: exact solutions, Sov. Phys. JETP
	62~(5) (1985) 894--899.
	
	\bibitem{akhmediev1987exact}
	N.~Akhmediev, V.~Eleonskii, N.~Kulagin, Exact first-order solutions of the
	nonlinear schr{\^o}dinger equation, Teoreticheskaya i Matematicheskaya Fizika
	72~(2) (1987) 183--196.
	
	\bibitem{akhmediev2009extreme}
	N.~Akhmediev, J.~M. Soto-Crespo, A.~Ankiewicz, Extreme waves that appear from
	nowhere: on the nature of rogue waves, Physics Letters A 373~(25) (2009)
	2137--2145.
	
	\bibitem{akhmediev2009rogue}
	N.~Akhmediev, A.~Ankiewicz, J.~M. Soto-Crespo, Rogue waves and rational
	solutions of the nonlinear Schr{\"o}dinger equation, Physical Review E 80~(2)
	(2009) 026601.
	
	\bibitem{ankiewicz2011rogue}
	A.~Ankiewicz, D.~J. Kedziora, N.~Akhmediev, Rogue wave triplets, Physics
	Letters A 375~(28-29) (2011) 2782--2785.
	
	\bibitem{kedziora2012second}
	D.~J. Kedziora, A.~Ankiewicz, N.~Akhmediev, Second-order nonlinear
	Schr{\"o}dinger equation breather solutions in the degenerate and rogue wave
	limits, Physical Review E 85~(6) (2012) 066601.
	
	\bibitem{gaillard2013degenerate}
	P.~Gaillard, Degenerate determinant representation of solutions of the
	nonlinear Schr{\"o}dinger equation, higher order peregrine breathers and
	multi-rogue waves, Journal of Mathematical Physics 54~(1) (2013) 013504.
	
	\bibitem{kedziora2013classifying}
	D.~J. Kedziora, A.~Ankiewicz, N.~Akhmediev, Classifying the hierarchy of
	nonlinear-Schr{\"o}dinger-equation rogue-wave solutions, Physical Review E
	88~(1) (2013) 013207.
	
	\bibitem{kruglov2005exact}
	V.~Kruglov, A.~Peacock, J.~Harvey, Exact solutions of the generalized nonlinear
	Schr{\"o}dinger equation with distributed coefficients, Physical Review E
	71~(5) (2005) 056619.
	
	\bibitem{yan2010nonautonomous}
	Z.~Yan, Nonautonomous “rogons” in the inhomogeneous nonlinear
	Schr{\"o}dinger equation with variable coefficients, Physics Letters A
	374~(4) (2010) 672--679.
	
	\bibitem{zhong2013rogue}
	W.-P. Zhong, M.~R. Beli{\'c}, T.~Huang, Rogue wave solutions to the generalized
	nonlinear Schr{\"o}dinger equation with variable coefficients, Physical
	Review E 87~(6) (2013) 065201.
	
	\bibitem{loomba2013optical}
	S.~Loomba, H.~Kaur, Optical rogue waves for the inhomogeneous generalized
	nonlinear schr{\"o}dinger equation, Physical Review E 88~(6) (2013) 062903.
	
	\bibitem{yang2018controllable}
	Z.~Yang, W.-P. Zhong, M.~Beli{\'c}, Y.~Zhang, Controllable optical rogue waves
	via nonlinearity management, Optics Express 26~(6) (2018) 7587--7597.
	
	\bibitem{zhong2014controllable}
	W.-P. Zhong, L.~Chen, M.~Beli{\'c}, N.~Petrovi{\'c}, Controllable
	parabolic-cylinder optical rogue wave, Physical Review E 90~(4) (2014)
	043201.
	
	\bibitem{li2018rogue}
	B.-Q. Li, Y.-L. Ma, Rogue waves for the optical fiber system with variable
	coefficients, Optik 158 (2018) 177--184.
	
	\bibitem{dai2012controllable}
	C.-Q. Dai, G.-Q. Zhou, J.-F. Zhang, et~al., Controllable optical rogue waves in
	the femtosecond regime, Physical Review E 85~(1) (2012) 016603.
	
	\bibitem{tiofack2015comb}
	C.~G.~L. Tiofack, S.~Coulibaly, M.~Taki, S.~De~Bi{\`e}vre, G.~Dujardin, Comb
	generation using multiple compression points of peregrine rogue waves in
	periodically modulated nonlinear Schr{\"o}dinger equations, Physical Review A
	92~(4) (2015) 043837.
	
	\bibitem{zhao2006gain}
	L.~Zhao, D.~Tang, T.~Cheng, C.~Lu, Gain-guided solitons in dispersion-managed
	fiber lasers with large net cavity dispersion, Optics Letters 31~(20) (2006)
	2957--2959.
	
	\bibitem{zhang2013experimental}
	L.~Zhang, A.~El-Damak, Y.~Feng, X.~Gu, Experimental and numerical studies of
	mode-locked fiber laser with large normal and anomalous dispersion, Optics
	Express 21~(10) (2013) 12014--12021.
	
	\bibitem{chandrasekhar2006performance}
	S.~Chandrasekhar, A.~Gnauck, Performance of mlse receiver in a
	dispersion-managed multispan experiment at 10.7 gb/s under nonlinear
	transmission, IEEE Photonics Technology Letters 18~(23) (2006) 2448--2450.
	
	\bibitem{torrengo2011experimental}
	E.~Torrengo, R.~Cigliutti, G.~Bosco, A.~Carena, V.~Curri, P.~Poggiolini,
	A.~Nespola, D.~Zeolla, F.~Forghieri, Experimental validation of an analytical
	model for nonlinear propagation in uncompensated optical links, Optics
	Express 19~(26) (2011) B790--B798.
	
	\bibitem{lin2011spin}
	Y.-J. Lin, K.~Jim{\'e}nez-Garc{\'\i}a, I.~B. Spielman, Spin--orbit-coupled
	Bose--Einstein condensates, Nature 471~(7336) (2011) 83--86.
	
	\bibitem{huang2016experimental}
	L.~Huang, Z.~Meng, P.~Wang, P.~Peng, S.-L. Zhang, L.~Chen, D.~Li, Q.~Zhou,
	J.~Zhang, Experimental realization of two-dimensional synthetic spin--orbit
	coupling in ultracold fermi gases, Nature Physics 12~(6) (2016) 540--544.
	
	\bibitem{matveev1991darboux}
	V.~B. Matveev, V.~Matveev, Darboux transformations and solitons, Springer-Verlag (1991).
	
	\bibitem{cieslinski2009algebraic}
	J.~L. Cie{\'s}li{\'n}ski, Algebraic construction of the Darboux matrix
	revisited, Journal of Physics A: Mathematical and Theoretical 42~(40) (2009)
	404003.
	
	\bibitem{akhmediev2011rogue}
	N.~Akhmediev, A.~Ankiewicz, J.~Soto-Crespo, J.~M. Dudley, Rogue wave early
	warning through spectral measurements?, Physics Letters A 375~(3) (2011)
	541--544.
	
	\bibitem{ling2013simple}
	L.~Ling, L.-C. Zhao, Simple determinant representation for rogue waves of the
	nonlinear Schr{\"o}dinger equation, Physical Review E 88~(4) (2013) 043201.
	
	\bibitem{chertkov2001pulse}
	M.~Chertkov, I.~Gabitov, J.~Moeser, Pulse confinement in optical fibers with
	random dispersion, Proceedings of the National Academy of Sciences 98~(25)
	(2001) 14208--14211.
	
	\bibitem{ablowitz2004dispersion}
	M.~J. Ablowitz, J.~T. Moeser, Dispersion management for randomly varying
	optical fibers, Optics letters 29~(8) (2004) 821--823.
	
	\bibitem{smith1996modulational}
	N.~Smith, N.~Doran, Modulational instabilities in fibers with periodic
	dispersion management, Optics letters 21~(8) (1996) 570--572.
	
	\bibitem{he2020dynamics}
	Y.~He, S.~Wang, A.~Yang, X.~Zeng, Dynamics of optical rogue wave generation in
	dispersion oscillating fibers, Optics Express 28~(14) (2020) 19877--19886.
	
	\bibitem{agrawal2013nonlinear}
	G.~P. Agrawal, Nonlinear Fiber Optics, Academic Press, 2013.
	
	\bibitem{biswas2010mathematical}
	A.~Biswas, D.~Milovic, M.~Edwards, Mathematical theory of dispersion-managed
	optical solitons, Springer Science \& Business Media, 2010.
	
	\bibitem{guo2012nonlinear}
	B.~Guo, L.~Ling, Q.~Liu, Nonlinear Schr{\"o}dinger equation: generalized
	Darboux transformation and rogue wave solutions, Physical Review E 85~(2)
	(2012) 026607.
	
	\bibitem{baronio2014vector}
	F.~Baronio, M.~Conforti, A.~Degasperis, S.~Lombardo, M.~Onorato, S.~Wabnitz,
	Vector rogue waves and baseband modulation instability in the defocusing
	regime, Physical Review Letters 113~(3) (2014) 034101.
	
	\bibitem{ablowitz2001long}
	M.~Ablowitz, J.~Hammack, D.~Henderson, C.~Schober, Long-time dynamics of the
	modulational instability of deep water waves, Physica D: Nonlinear Phenomena
	152 (2001) 416--433.
	
	\bibitem{van2001}
	G.~Van~Simaeys, P.~Emplit, M.~Haelterman, Experimental demonstration of the
	Fermi-Pasta-Ulam recurrence in a modulationally unstable optical wave,
	Physical review letters 87~(3) (2001) 033902.
	
	\bibitem{govindarajan2017}
	A.~Govindarajan, B.~A. Malomed, A.~Mahalingam, A.~Uthayakumar, Modulational
	instability in linearly coupled asymmetric dual-core fibers, Applied Sciences
	7~(7) (2017) 645.
	
	\bibitem{randoux2016inverse}
	S.~Randoux, P.~Suret, G.~El, Inverse scattering transform analysis of rogue
	waves using local periodization procedure, Scientific reports 6~(1) (2016)
	1--11.
	
	\bibitem{dudley2014}
	J.~M. Dudley, F.~Dias, M.~Erkintalo, G.~Genty, Instabilities, breathers and
	rogue waves in optics, Nature Photonics 8~(10) (2014) 755--764.
	
	\bibitem{ablowitz1996computational}
	M.~Ablowitz, B.~Herbst, C.~Schober, Computational chaos in the nonlinear
	Schr{\"o}dinger equation without homoclinic crossings, Physica A: Statistical
	Mechanics and its Applications 228~(1-4) (1996) 212--235.
	
	\bibitem{grinevich2018finite}
	P.~Grinevich, P.~Santini, The finite gap method and the analytic description of
	the exact rogue wave recurrence in the periodic NLS cauchy problem. 1,
	Nonlinearity 31~(11) (2018) 5258.
	
	\bibitem{randoux2016}
	S.~Randoux, P.~Suret, G.~El, Inverse scattering transform analysis of rogue
	waves using local periodization procedure, Scientific Reports 6~(1) (2016)
	1--11.
	
\end{thebibliography}
\end{document}